%% file: main.tex
\documentclass[sigconf,screen]{acmart}
\PassOptionsToPackage{bookmarks=false}{hyperref}

\AtBeginDocument{%
  \providecommand\BibTeX{{%
    \normalfont B\kern-0.5em{\scshape i\kern-0.25em b}\kern-0.8em\TeX}}}

\setcopyright{acmcopyright}
\acmPrice{15.00}
\acmDOI{10.1145/xxxxxxx.xxxxxxx}
\acmYear{2024}
\copyrightyear{2024}
\acmConference[ICSE 2024]{46th International Conference on Software Engineering}{April 2024}{Lisbon, Portugal}

\usepackage[ruled,vlined,lined,commentsnumbered]{algorithm2e}
\usepackage{booktabs}
\usepackage{moreverb}
\usepackage{fontenc}
\usepackage{amsmath}

\usepackage{amssymb}
\usepackage{fancybox}
\usepackage{color}
\usepackage{colortbl}
\usepackage{array}
\usepackage{diagbox}
\usepackage{flushend}
\usepackage{multirow}
\usepackage{multicol}
\usepackage{makecell}
\usepackage{listings}
\usepackage{makecell}
\usepackage{graphicx}
\usepackage{setspace}
\usepackage{soul}
\usepackage{listings}
\usepackage{csvsimple}
\usepackage{longtable}
\usepackage[skip=0pt,font=small,labelfont=bf]{caption}
\usepackage{url}
\usepackage{lscape}
\usepackage{rotating}
\usepackage{tikz}
\usepackage{enumitem}
\usepackage{subcaption}
\usepackage{wrapfig}
\usepackage[most]{tcolorbox}
\makeatletter
\@namedef{ver@lineno.sty}{9999/12/31}
\@namedef{opt@lineno.sty}{}
\makeatother
\usepackage{threeparttable}
\usepackage{float}
\usepackage{marvosym}
\usepackage{pifont}
\usepackage{xcolor}

\newcolumntype{L}[1]{>{\raggedright\let\newline\\\arraybackslash\hspace{0pt}}m{#1}}
\newcolumntype{C}[1]{>{\centering\let\newline\\\arraybackslash\hspace{0pt}}m{#1}}
\newcolumntype{R}[1]{>{\raggedleft\let\newline\\\arraybackslash\hspace{0pt}}m{#1}}

\definecolor{codegreen}{rgb}{0,0.6,0}
\definecolor{codegray}{rgb}{0.5,0.5,0.5}
\definecolor{codepurple}{rgb}{0.58,0,0.82}
\definecolor{backcolour}{rgb}{0.95,0.95,0.92}

\makeatletter
\def\UrlAlphabet{%
      \do\a\do\b\do\c\do\d\do\e\do\f\do\g\do\h\do\i\do\j%
      \do\k\do\l\do\m\do\n\do\o\do\p\do\q\do\r\do\s\do\t%
      \do\u\do\v\do\w\do\x\do\y\do\z\do\A\do\B\do\C\do\D%
      \do\E\do\F\do\G\do\H\do\I\do\J\do\K\do\L\do\M\do\N%
      \do\O\do\P\do\Q\do\R\do\S\do\T\do\U\do\V\do\W\do\X%
      \do\Y\do\Z}
\def\UrlDigits{\do\1\do\2\do\3\do\4\do\5\do\6\do\7\do\8\do\9\do\0}

\g@addto@macro{\UrlBreaks}{\UrlOrds}
\g@addto@macro{\UrlBreaks}{\UrlAlphabet}
\g@addto@macro{\UrlBreaks}{\UrlDigits}
\makeatother

\lstdefinestyle{mystyle}{
    backgroundcolor=\color{white},
    commentstyle=\color{codegreen},
    keywordstyle=\color{magenta},
    numberstyle=\tiny\color{codegray},
    stringstyle=\color{codepurple},
    basicstyle=\footnotesize,
    breakatwhitespace=false,
    breaklines=true,
    captionpos=b,
    keepspaces=true,
    showspaces=false,
    showstringspaces=false,
    showtabs=false,
    tabsize=2
}

\setlist{noitemsep} 

\definecolor{darkpastelred}{rgb}{0.76, 0.23, 0.13}
\definecolor{ao(english)}{rgb}{0.0, 0.5, 0.0}

\definecolor{darkpastelred}{rgb}{0.76, 0.23, 0.13}
\definecolor{ao(english)}{rgb}{0.0, 0.5, 0.0}

\lstdefinelanguage{diff}{
  morecomment=[f][\color{blue}]{@@},     
  morecomment=[f][\color{red}]-,         
  morecomment=[f][\color{codegreen}]+,       
  morecomment=[f][\color{red}]{---}, 
  morecomment=[f][\color{codegreen}]{+++},
}

\definecolor{yellow}{RGB}{255,255,153}
\definecolor{grey}{RGB}{224,224,224}
\definecolor{lightgreen}{HTML}{99d8c9}

\newboolean{showcomments}
\setboolean{showcomments}{true}
\ifthenelse{\boolean{showcomments}}
 { \newcommand{\mynote}[2]{
      \fbox{\bfseries\sffamily\scriptsize#1}
        {\small$\blacktriangleright$\textsf{\emph{#2}}$\blacktriangleleft$}}}
        { \newcommand{\mynote}[2]{}}

\definecolor{DarkOrange}{rgb}{0.8,0.3,0.0}
\definecolor{DarkCyan}{rgb}{0.0, 0.55, 0.55}
\definecolor{amber}{rgb}{1.0, 0.75, 0.0}

\newcommand{\etal}{\emph{et~al.}\xspace}

\newcolumntype{?}{!{\vrule width 1pt}}

\newcommand*{\ie}{i.e., }
\newcommand*{\eg}{e.g., }

\newcommand*{\mycode}{\fontfamily{lmtt}\selectfont}

\newcommand{\find}[1]{
\begin{tcolorbox}[leftrule=1mm,rightrule=1mm,toprule=0mm,bottomrule=0mm,left=1pt,right=1pt,top=0.5pt,bottom=0.5pt]
\em #1
\end{tcolorbox}
}

\begin{document}

\title{Large Language Models are Few-Shot Summarizers:\\
Multi-Intent Comment Generation via In-Context Learning
}

\thanks{$^\dagger$ Shangwen Wang and Dezun Dong are the corresponding authors. \\
Shangwen Wang and Xiaoguang Mao are with the Key Laboratory of Software Engineering for Complex Systems.\\
This work is supported by the National Key Research and Development Program Project ``Heterogeneous Computing Fusion of Cross-Domain Resources'' No.2022YFB4501702.
}


\author{Mingyang Geng}
\email{gengmingyang13@nudt.edu.cn}
\affiliation{%
  \institution{College of Computer Science, National University of Defense Technology}
 	\country{Changsha, China}
}

\author{Shangwen Wang}
\email{wangshangwen13@nudt.edu.cn}
\affiliation{%
  \institution{College of Computer Science, National University of Defense Technology}
 	\country{Changsha, China}
}

\author{Dezun Dong}
\email{dong@nudt.edu.cn}
\affiliation{%
  \institution{College of Computer Science, National University of Defense Technology}
 	\country{Changsha, China}
}

\author{Haotian Wang}
\email{wanghaotian13@nudt.edu.cn}
\affiliation{%
  \institution{College of Computer Science, National University of Defense Technology}
 	\country{Changsha, China}
}

\author{Ge Li}
\email{lige@pku.edu.cn}
\affiliation{%
  \institution{Key Lab of High Confidence Software Technology, Peking University}
 	\country{Beijing, China}
}

\author{Zhi Jin}
\email{zhijin@pku.edu.cn}
\affiliation{%
  \institution{Key Lab of High Confidence Software Technology, Peking University}
 	\country{Beijing, China}
}

\author{Xiaoguang Mao}
\email{xgmao@nudt.edu.cn}
\affiliation{%
  \institution{College of Computer Science, National University of Defense Technology}
 	\country{Changsha, China}
}

\author{Xiangke Liao}
\email{xkliao@nudt.edu.cn}
\affiliation{%
  \institution{College of Computer Science, National University of Defense Technology}
 	\country{Changsha, China}
}






\input{0.abstract}



\maketitle

\input{1.intro}

\input{2.background}

\input{3.setup}

\input{4.evaluation}

\input{6.discussion}

\input{7.relatedwork}

\input{8.conclusion}


\balance




\end{document}

%% file: 0.abstract.tex
\begin{abstract} 

Code comment generation aims at generating natural language descriptions for a code snippet to facilitate developers' program comprehension activities.
Despite being studied for a long time, a bottleneck for existing approaches is that given a code snippet, they can only generate one comment while developers usually need to know information from diverse perspectives such as what is the functionality of this code snippet and how to use it.
To tackle this limitation, this study empirically investigates the feasibility of utilizing large language models (LLMs) to generate comments that can fulfill developers' diverse intents.
Our intuition is based on the facts that (1) the code and its pairwise comment are used during the pre-training process of LLMs to build the semantic connection between the natural language and programming language, and (2) comments in the real-world projects, which are collected for the pre-training, usually contain different developers' intents.
We thus postulate that the LLMs can already understand the code from different perspectives after the pre-training.
Indeed, experiments on two large-scale datasets demonstrate the rationale of our insights: by adopting the in-context learning paradigm and giving adequate prompts to the LLM (\eg providing it with ten or more examples), the LLM can significantly outperform a state-of-the-art supervised learning approach on generating comments with multiple intents.
Results also show that customized strategies for constructing the prompts and post-processing strategies for reranking the results can both boost the LLM's performances, which shed light on future research directions for using LLMs to achieve comment generation.

\end{abstract}

%% file: 1.intro.tex
\section{Introduction}
\label{sec:intro}

Code comment generation (a.k.a. code summarization) targets the ambition of automatically generating a concise and fluent natural language description of source code.
It is considered as a critical way to facilitate program comprehension since developers usually forget or have no time to write such comments, and thus holds the potential of boosting software development and maintenance activities.
During the years, a number of studies have been devoted into advancing the state of the art in this domain \cite{iyer2016summarizing,alon2018code2seq,hu2018deep}.
For instance, information retrieval techniques, which focus on extracting some important tokens from the code, are used in the early stage \cite{haiduc2010use,rodeghero2014improving}, followed by some recent works applying advanced deep learning techniques on this task, such as the neural machine translation (NMT) model \cite{hu2018deep,alon2019code2seq}.

Despite the achieved tremendous progress in this domain, one critical problem that downgrades the practicality of existing code comment generation approaches is that they can only generate comments describing one aspect of a given code snippet (and thus a one-to-one mapping).
In practice, however, developers often write comments with diverse intents to summarize the code from different perspectives (\eg what is the main functionality of the code and how can we use it). 
For instance, Zhai \etal \cite{zhai2020cpc} manually checked comments from real-world projects and identified six categories of intents hidden in the comments (as shown in Table~\ref{tab:multi-intent}).
Mu \etal \cite{mu2023developer} did the statistics of top-starred Java projects on GitHub and found that around 67\% of the methods contain more than one intent in their comments.
The above observations indicate that what developers really need is a one-to-many mapping (\ie generating multiple comments that summarize the given code from different perspectives), which is referred to as the {\bf multi-intent comment generation} task in this paper.

\input{tables/intent_taxonomy}

To tackle the aforementioned task, Mu \etal \cite{mu2023developer} proposed an approach named DOME, where an attention mechanism is used to focus on different parts of code for different intents. 
However, DOME is based on supervised learning, which limits its effectiveness due to the amount of data available for training. 
To address the data shortage problem, we propose to borrow the weapon of large language models (LLMs) \cite{brown2020language}, which are pre-trained on a data corpus of a very large scale in the self-supervised manner and have captured a lot of domain knowledge during such a process.
The application of LLMs to the multi-intent comment generation task is motivated by two factors. 
Firstly, LLMs designed for the code domain are typically pre-trained using code and its associated pairwise comments to establish semantic connections between programming language and natural language \cite{feng2020codebert,wang2021codet5}. 
For example, the commonly used pre-training task, masked language modeling \cite{feng2020codebert,guo2021graphcodebert,devlin2019bert}, is specifically intended to align programming language and natural language representations. 
Secondly, existing research has shown that code comments from real-world projects, which form the training corpus for LLMs, often contain multiple intents \cite{mu2023developer}. 
As a result, during pre-training, LLMs are trained to understand code from various perspectives, potentially allowing them to capture different code semantics. 
Thus, by fully exploiting the capabilities of pre-trained LLMs, we can achieve good performances on the multi-intent comment generation task.


Recently, in-context learning has been shown to be an effective way to exploit the domain knowledge hidden in the LLMs \cite{brown2020language,chen2021evaluating,rubin2021learning,nashid2023retrieval}, since the format of the inputs to the model can be consistent to that during the pre-training process. 
Inspired by these studies, we aim to investigate the feasibility of addressing the multi-intent comment generation task with in-context learning.
Generally, in-context learning requires to provide a prompt to the model which is composed of a natural language instruction describing the detailed information of the task, (optionally) a handful of examples demonstrating how the task could be well done, and a query that is required to be addressed.
Therefore, a follow-up question is that, with in-context learning, how can we obtain better results from the LLMs (\eg if it is possible by designing prompts that can guide the LLMs towards the desired output).
To provide empirical evidence on the aforementioned questions, we investigate the following aspects in this study:
(a) Can the LLMs support to accomplish the multi-intent comment generation task using the in-context learning paradigm?
(b) Can we improve the performance of the LLMs by designing customized demonstration selection strategies? 
and (c) Can we improve the performance of the LLMs by designing customized strategies to post-process the obtained results?



To that end, we perform extensive experiments on two large-scale Java language datasets, which are Funcom \cite{leclair2019neural} and TLC \cite{hu2018summarizing}.
We use the OpenAI Codex model as the representative LLM because of its superior performances on several code intelligence tasks \cite{prenner2022can,nashid2023retrieval}.
Our study makes the following important findings:
\begin{itemize}
    
    \item [F1:] When the LLM is not adequately prompted (\ie the number of demonstration examples is less than 10), the potential of the LLMs may not be fully exploited and the effectiveness is sub-optimal compared with that of the state-of-the-art supervised learning approach, DOME;
    in contrast, when the number of demonstration examples reaches ten, the LLM is more adequately prompted and its performance exceeds that of the DOME approach.
    
    

    \item[F2:] Demonstration selection strategies can help LLMs better understand the on-going task and thus enhance their effectiveness to a large extent: when the number of examples is ten and the code snippets which are most similar to the target one are used as the demonstration examples, the BLEU values of Codex can be increased by 97\% and 131\% on the two datasets, respectively, compared with random selection.
    
    \item[F3:] The outputs of LLMs can be reranked based on simple heuristics to achieve further performance enhancement: compared with the experiment setting mentioned above, the BLEU values of Codex can be improved by 9.9\% and 9.6\%, respectively, on the two datasets if the comment of the corpus code which is similar to the target one can be used for guiding the output reranking. 
    
\end{itemize}

Our study demonstrates that LLMs can potentially be applied to multi-intent comment generation since it builds strong performance baselines on this task, which should be considered by tool designers in future evaluation. 
Further implications include that devising better demonstration selection strategies as well as reranking strategies are both promising research directions.

%% file: tables/intent_taxonomy.tex
\begin{table*}[!t]
\caption{The intent taxonomy of code comments \cite{chen2021my,zhai2020cpc}.}
\centering
\resizebox{0.75\linewidth}{!}{
\begin{tabular}{c|l|l}
\toprule
{\bf Category} & {\bf Definition} & {\bf Example} \\
\hline
What&Describes the functionality of
a method & \makecell[l]{``Checks if the tile units at the given coordinates\\ are displayed on the screen''}\\
\hline
Why&\makecell[l]{Explains the reason why a
method is provided \\or the
design rationale of the method} & \makecell[l]{``Prepare to start making calls to the currently\\ registered callbacks''}\\
\hline
How-to-use&\makecell[l]{Describes the usage or the
expected set-up of\\ using a
method}&``Code executed before the intercepted method'' \\
\hline
How-it-is-done&Describes the implementation
details of a method&\makecell[l]{``Ends the current table, discards it and pops the\\ top of the stack to be the new current table''}\\
\hline
Property&\makecell[l]{Asserts properties of a method
including\\ pre-conditions or
post-conditions of a method}&\makecell[l]{``Returns true if the value is a string that matches\\ a regex''}\\
\hline
Others&Unspecified or ambiguous comments&``I am done with the model, free the resources ''\\
\bottomrule
\end{tabular}
}
\label{tab:multi-intent}
\end{table*}

%% file: 2.background.tex
\section{Background and Related Works}
\label{sec:bg}

\subsection{Comment Generation}


Automatic code comment generation, which aims at summarizing code with concise natural language descriptions, is a critical task to facilitate program comprehension. 
Many approaches have been proposed to construct a set of manually-defined complex rules, based on which comments can be generated following specific templates \cite{haiduc2010use,hill2009automatically}.
With the recent advancement of the deep learning, a hot line of researches has suggested applying deep neural networks (DNNs) to this task. 
By modeling code as the input and comment as the output, such neural comment generation (NCG) approaches automatically learn a function, which is usually a DNN model such as the neural machine translation model,
that can produce the output given the input. 
Such a DNN model is learned using existing large-scale code-comment pairwise data. 
CodeNN \cite{iyer2016summarizing} is an early attempt in this direction that uses only code token sequences, followed by various approaches that utilize the AST structure \cite{alon2018code2seq,hu2018deep,hu2020deep},
API knowledge \cite{hu2018summarizing}, type information \cite{cai2020tag}, global context \cite{ bansal2021project,haque2020improved, wang2021cocosum}, reinforcement learning \cite{ wan2018improving, wang2020reinforcement, geng2023interpretation}, multi-task learning \cite{xie2021exploiting}, dual learning \cite{wei2019code,ye2020leveraging}, pre-trained language models \cite{feng2020codebert,wang2021codet5,geng2022fine}, and hybrid approaches \cite{wei2020retrieve,zhang2020retrieval}. 
In addition, a number of works also focus on generating latest and informative comments based on outdated comments (a.k.a comment updating) \cite{lin2021automated,lin2022predictive}.

The aforementioned approaches, however, can only generate comments describing one aspect of a given code snippet, which limits their practicality since developers usually express multiple intents when commenting the code
\cite{zhai2020cpc,chen2021my,mu2023developer}.
That is to say, merely generating comments describing a specific aspect of a code snippet (\eg the functionality of the code) may not meet the developers' requirements about comprehensively summarizing the code (\eg how to use the code).
Specifically, according to the previous studies \cite{zhai2020cpc,chen2021my,mu2023developer}, developers usually have six categories of intents when commenting the code, \ie {\em what}, {\em why}, {\em how-to-use}, {\em how-it-is-done}, {\em property}, and {\em others}. 
In Table \ref{tab:multi-intent}, we list the detailed definition and example for each category.
The fact that developers usually express multiple intents in the comments cast threats to the practicality of existing single-intent comment generation techniques.
To address this challenge, Mu \etal \cite{mu2023developer} propose a developer-intent driven code comment generation approach DOME, which aims to produce a comment coherent with a given intent.
It works by leveraging the attention mechanism guided by the given intent to focus on the most relevant information from the code.
To our best knowledge, DOME is so far the only existing technique that can generate diverse comments given different categories of intents.



\subsection{Large Language Models}


Large language models (LLMs) trained on massive corpora of unlabelled data have been shown to perform well on a wide range of tasks, including natural language generation, semantic parsing, and code generation \cite{brown2020language,devlin2018bert,raffel2020exploring}.
The reason for their strong power can be concluded as they do not need task-specific training data and can be pre-trained on tremendous in-the-wild data in a self-supervised manner (a.k.a. pre-training), so that sufficient domain knowledge can be captured.
The pioneer of this direction, the GPT model \cite{radford2018improving}, was firstly proposed in 2018. 
After that, a number of follow-up studies continuously enhance the state-of-the-art performances by adjusting the model architecture (\eg BERT \cite{devlin2018bert}) or increasing the total amount of parameters (\eg GPT-3 \cite{brown2020language}).



Codex, released by OpenAI, is an LLM based on the GPT-3 architecture (\ie contains a Transformer-based decoder) \cite{ref:codex}.
It powers GitHub Copilot, an AI pair programmer that generates the whole code function given a natural language description.
Codex is trained on a massive code corpus containing code-comment pairwise examples from many programming languages including Python, JavaScript, C/C++, Go, Perl, PHP, Ruby, Swift, TypeScript, SQL and Shell.
Similar to GPT-3, Codex adopts the auto-regressive manner during the pre-training, in which given a sequence of code/comment tokens, it is trained to predict the next token and the predicted token is recursively used as the input for the next prediction until the end of the sequence.
In our study, we use Codex as the representative LLM since it is a popular LLM in the software engineering domain and has been widely studied in the literature \cite{dakhel2022github,prenner2022can,kolak2022patch,fan2022improving,pearce2022asleep,chen2022codet,zhang2022coder,ni2023lever}.


\subsection{In-Context Learning}


Previously, to apply a pre-trained model on downstream tasks, users need to further train it on the labelled data of downstream tasks in a supervised manner (a.k.a. fine-tuning) \cite{devlin2018bert,liu2016recurrent}.
Compared with training a model from scratch, this paradigm can exploit the knowledge learned by the pre-trained model and thus achieve better performance \cite{mastropaolo2022using,li2022auger}.
Such a paradigm, however, mainly has two limitations. First, the data used for pre-training and fine-tuning are in different formats, which makes the learned knowledge of the model cannot be fully leveraged during the fine-tuning process \cite{wang2022no}.
Second, the fine-tuning process can be extremely time-consuming and resource-intensive, especially when it comes to large language models which usually contain billions of parameters \cite{brown2020language}.



To address the aforementioned limitations, {\bf in-context learning} is recently proposed and quickly becomes a research hotspot after that \cite{brown2020language}.
Such a paradigm denotes that a few training examples and/or task descriptions together with a developer query that needs to be answered are sent into a large language model to produce a response of the query, without any parameter update. 
Basically, in the in-context learning paradigm, a prompt needs to be provided for a code intelligence task, \eg code summarization.
By employing prompts, large language models are shown to be effective in different tasks that the model is not explicitly trained on, without the need of task-specific data \cite{wang2022no}.


Generally, the rationale of the in-context learning is that since large language models have been trained on corpora of a very large scale, they must have absorbed much domain knowledge and are thus expected to generalize well to unseen tasks without fine-tuning \cite{brown2020language}. 
Our study shares a similar motivation. 
Specifically, considering that (1) large language models, \eg Codex, are trained on a large-scale corpus containing tremendous amount of code-comment pairwise data from real-world, and (2) the real-world comments usually contain different categories of developers' intents, we postulate that the large language models are capable of understanding the code from different perspectives and thus hold the potential to generate comments with diverse intents given a code snippet.
By using the in-context learning, such potentials of LLMs can be exploited.

%% file: 3.setup.tex
\section{Study Design}
\label{sec:exp}

\begin{figure*}[!t]
     \centering
     \includegraphics[width=0.65\textwidth]{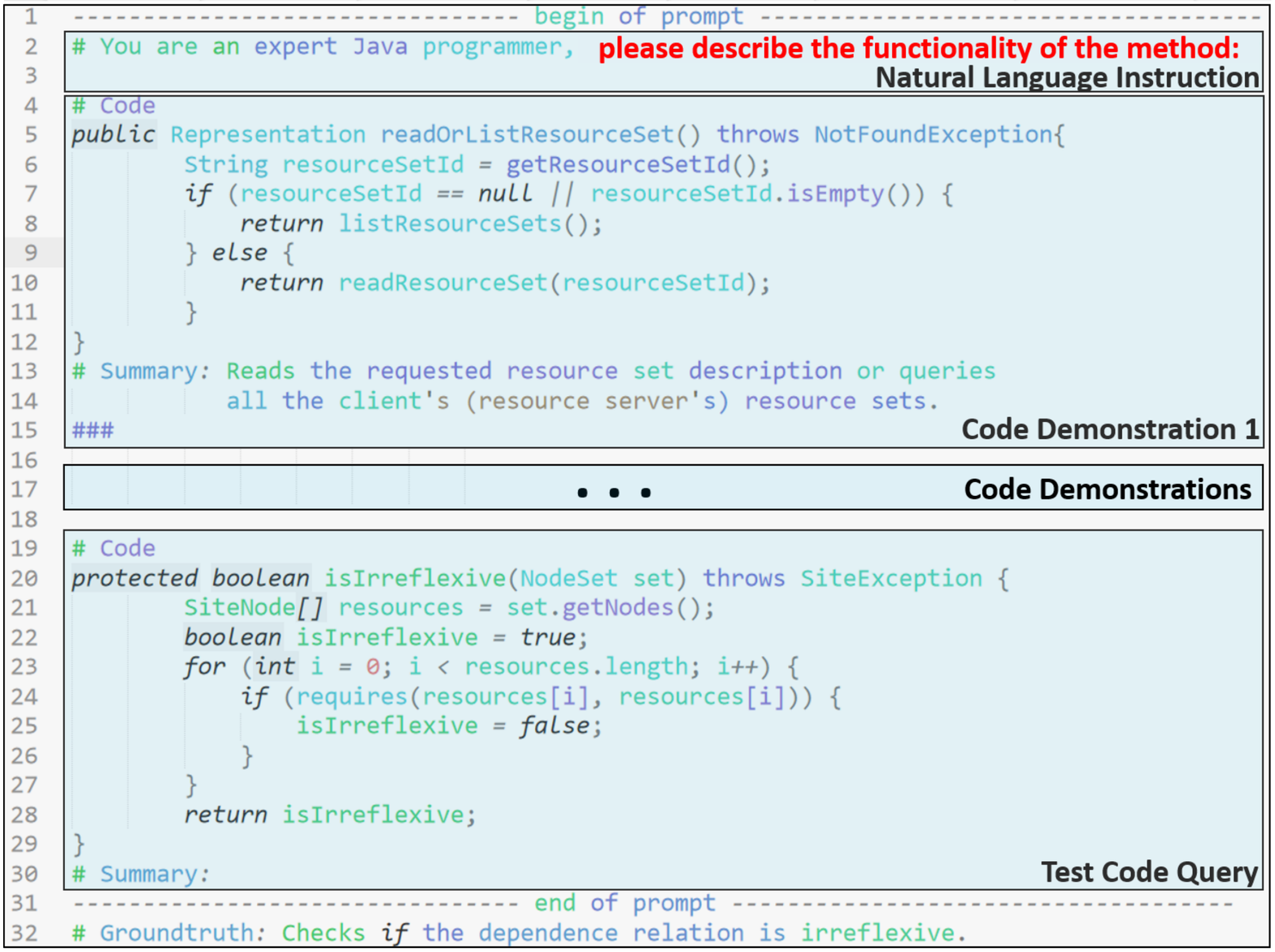}
     \caption{Multi-intent code summarization prompt template.}
     \label{fig:prompt}
 \end{figure*}

\subsection{Research Questions}

The goal of our study is to investigate the effectiveness of large language models on multi-intent comment generation using the in-context learning paradigm.
To this end, we propose to answer the following research questions.

\begin{itemize}[leftmargin=*]
    \item {\bf RQ1: What is the effectiveness of Codex on multi-intent comment generation using zero-shot, one-shot, and few-shot learning?}
    As the very first RQ, we aim at investigating the feasibility of addressing the multi-intent comment generation problem with in-context learning. Specifically, we do not use any customized design and only select code demonstrations randomly. Our target is to investigate how effective is the vanilla in-context learning compared with the state-of-the-art DOME approach. The results can also reflect to what extent the number of demonstrations (\ie zero-shot, one-shot, and few-shot) affect the effectiveness.

    \item {\bf RQ2: Can the effectiveness be improved by retrieval-based demonstration selections?}
    Some recent works have demonstrated that the quality of the demonstrations in the prompt can significantly impact the effectiveness of in-context learning \cite{nashid2023retrieval,rubin2021learning,min2022rethinking}.
    Inspired by these studies, we propose to investigate whether customized demonstration selection approaches can help improve the model's performance. Specifically, to answer this question, we design two retrieval-based approaches that select code examples similar to the code specified in the developer query, and evaluate their effectiveness.

    \item {\bf RQ3: Can the effectiveness be improved by reranking strategies?}
    A large language model experiences a sampling process to obtain the outputs \cite{zhang2022coder,ni2023lever,chen2021evaluating,shi2022natural}. 
    That is to say, a developer can obtain different results from the model for the identical input. 
    In this RQ, we further investigate the feasibility of boosting the model's performance in a post-processing manner: by first obtaining a number of results and then reranking them through a pre-defined heuristic.
    Answering such a question can provide guidance for applying the approach in practice: it can make us clear about 
    to what extent we can obtain more qualified results by sampling multiple outputs.
    
\end{itemize}


\subsection{The Prompt Template for Multi-Intent Comment Generation}

Formally, a prompt is defined as $P = \left\{x_{\text {test }}+\mathcal{C} \mathcal{D}+\mathcal{N} \mathcal{L}\right\}$, where $\mathcal{N} \mathcal{L}$ is a natural language template, $\mathcal{C} \mathcal{D}=$ $\left\{\left(x_i, y_i\right)\right\}_{i=1}^n$ is a set of code demonstrations composed by input code sequence $(x_{i})$ and desired output sequence $(y_{i})$,
and ${x_{\text {test }}}$ is a developer query to be inferred.
Specifically, if $i=0$ which means there is no code demonstration, the setting is known as {\em zero-shot learning};
if $i=1$ which means there is only one code demonstration, the setting is known as {\em one-shot learning};
and {\em few-shot learning} means there is a number of code demonstrations.
Also, there is a constraint that $\operatorname{size}(\mathcal{P}) \leq$ context-window, which means the prompt should fit within the context window limit of the language model. \footnote{Language models limit the amount of contextual information that could be fed it to the model; the context window for Codex is limited to 8,000 tokens}

Figure~\ref{fig:prompt} illustrates a prompt template for the multi-intent comment generation task. 
The input prompt contains two sections: the code demonstrations $\mathcal{C} \mathcal{D}$ and the query  ${x_{\text {test }}}$.
The natural language instructions are denoted by the lines starting with the special token ``\#''.
In the first line of the prompt, we first tell the model the specific programming language it is working on (\eg Java) and then the desired intent of the comment, as highlighted in the red, is specified by following the definitions shown in Table~\ref{tab:multi-intent}.
In concrete, for the ``what'' intent, we add the prompt ``Describe the functionality of the method''; for the ``why'' intent, we add the prompt ``Explain the reason why the method is provided or the design rationale of the method''; for the ``how-to-use'' intent, we add the prompt ``Describe the usage or the expected set-up of using the method''; for the ``how-it-is-done'' intent, we add the prompt ``Describe the implementation details of the method''; for the ``property'' intent, we add the prompt ``Assert properties of the method including pre-conditions pr post-conditions of the method''.
In this example, the illustrated prompt aims at generating a comment that fulfills the ``what'' intent.
The first line is then followed by a number of code demonstrations that can help the LLM to understand the expected behavior and each demonstration contains one code snippet and one corresponding comment within the desired intent category.
Each code demonstration is separated with a delimiter ``\#\#\#''.
Finally, the model is asked to output the desired comment of the query code, which is shown at the bottom of the figure.

\begin{figure*}[!t]
     \centering
     \includegraphics[width=0.65\textwidth]{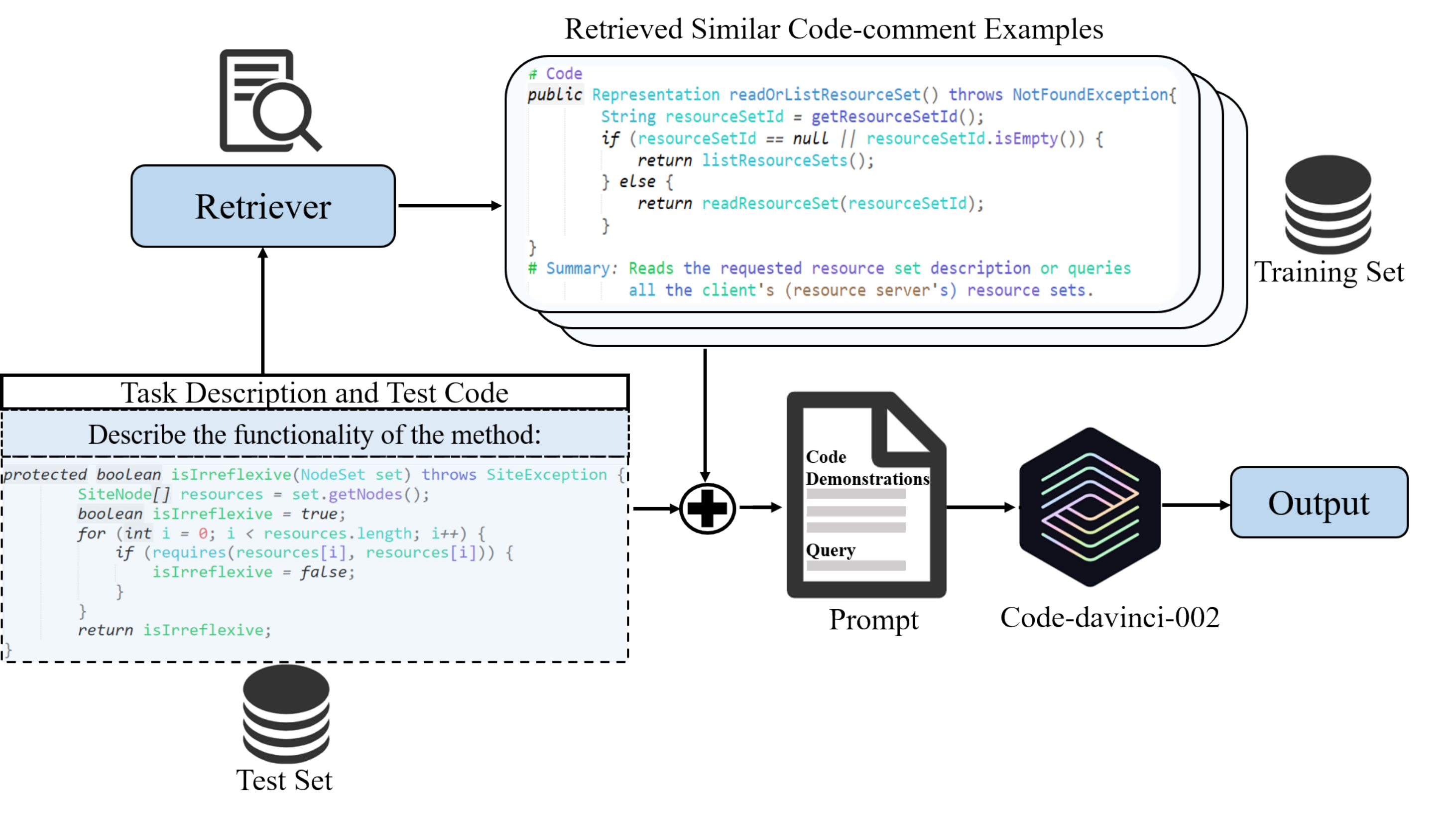}
     \caption{Overview of our in-context learning-based code summarization.}
     \label{fig:framework}
 \end{figure*}

\subsection{Demonstration Retrieval}

Note that the code demonstrations used in RQ1 are randomly selected from a corpus. While in RQ2, we aim at investigating whether customized demonstration selection can enhance the effectiveness. Therefore, we design two strategies to retrieve similar code demonstration examples from the corpus whose comments' intents belong to the desired category. 
The rationale is that a few demonstrations that are similar to the target one may help the model better understand the desired behaviour \cite{nashid2023retrieval,rubin2021learning,min2022rethinking}. 
The whole process of such a paradigm is shown in Figure~\ref{fig:framework}: given a code snippet and the required intent category, we select code examples that are similar to the target one and use the retrieved code together with their comments to construct a prompt whose template is shown in Figure~\ref{fig:prompt}.
The prompt is used to query the model and obtain the results.
We next introduce the two retrieval strategies in detail.

\begin{itemize}[leftmargin=*]
    \item \textbf{Token-based:} 
    The most commonly-used strategy to identify similar code is focusing on the overlap with respect to the code tokens \cite{kamiya2002ccfinder,zhang2023efficient,golubev2021multi}. 
    Inspired by these studies, our first retrieval strategy is also based on the token level information, \ie to rank the code snippets from the code corpus based on their token similarities with the target code.
    In concrete, we first pre-process the target code snippet and the code snippets in the retrieved code corpus by removing the keywords defined in the programming language (\ie Java in our study). The behind intuition is that such frequently-used tokens may bring side effects to the similarity calculation because a large number of code snippets would contain them, inspired by the recent study \cite{eghbali2022crystalbleu}.
    Then, we further split identifiers into sub-tokens to adequately leverage the semantic information hidden in the identifier names \cite{pradel2018deepbugs}.
    Specifically, such a process is achieved by utilizing the camel cases and the underscore naming convention of Java language.
    Finally, we convert all the sub-tokens to lower case.
    As for the token-based similarity between a candidate code snippet and the target code ($s_{token}$), we exploit the Jaccard Coefficient \cite{niwattanakul2013using} for the calculation, which is defined as follows:
       $ s_{\text {token }}=\frac{\mid \text { tokens }_{\text {target }} \cap \text { tokens }_{\text {candidate }} \mid}{\mid \text { tokens }_{\text {target }} \cup \text { tokens }_{\text {candidate }} \mid} $
    where $tokens_{target}$ denotes the sub-token list of the target code and $tokens_{candidate}$ denotes the sub-token list of the candidate code.
    The value of $s_{token}$ ranges from 0 to 1. A larger value of $s_{token}$ indicates a higher similarity between the target code and the candidate code from the retrieved set.
    
    \item \textbf{Semantic-based:} 
    Recent studies in the clone detection domain have also revealed that beyond the lexical level code token similarity, understanding the code semantics is also important for finding similar code \cite{zeng2022degraphcs,wang2020detecting}. 
    Therefore, our second strategy relies on the code semantics to retrieve similar code snippets. 
    Specifically, we exploit the pre-trained sentence transformer model \cite{reimers2019sentence}, which has been demonstrated to be capable of accurately capturing the semantics of code snippets by a recent study \cite{nashid2023retrieval}, to encode the code snippets as vectors which contain the corresponding semantic information. \footnote{We employ the st-codesearch-distilroberta-base model released at \url{https://huggingface.co/flax-sentence-embeddings/st-codesearch-distilroberta-base}, which was pre-trained on the CodeSearchNet dataset \cite{husain2019codesearchnet}}
    The cosine similarity is exploited to retrieve the similar candidate code snippets whose vectors are close to that of the target code snippet in the vector space. 

\end{itemize}

\subsection{Reranking Strategy}

To rerank the generated comments, our intuition is that similar code snippets usually share similar comments, which is a common sense in the literature \cite{wong2013autocomment,wong2015clocom,wei2020retrieve,li2021editsum}. 
Therefore, our strategy is to rerank the generated comments based on their similarities to the comment of the code snippet in the retrieval corpus that is similar to the target code.
Specifically, we use the comment of the code snippet that is the most similar to the target code as the reference and also calculate comment similarities from two perspectives, \ie the token-based and the semantic-based.
For the {\bf token-based} strategy, we focus on the token level information, since tokens in the comments are usually natural language words that have clear semantics. 
For the {\bf semantic-based}, we exploit again the pre-trained sentence transformer model \cite{reimers2019sentence}, embed the whole comment into a semantic vector, and calculate the cosine similarities.

\subsection{Datasets}

In this study, we use the multi-intent comment generation datasets released by the previous study \cite{mu2023developer} as our evaluation datasets. 
In concrete, we use two datasets of Java programming language, \ie the Funcom \cite{leclair2019neural} and TLC \cite{hu2018summarizing} datasets, both of which are the most widely-used datasets for the code comment generation task \cite{zhang2020retrieval, ahmad2020transformer,cheng2021gn,gao2023code,leclair2021ensemble}.
Funcom contains 2.1M code-comment pairs from 29K Java projects, which were collected by Lopes \etal \cite{ref:funcom} and further cleaned by LeClair \etal \cite{leclair2019neural}. 
TLC contains 87,136 code-comment pairs collected from more than 9K Java projects created from 2015 to 2016 with at least 20 stars.
The intent categories of each comment in these two datasets are labelled by Mu \etal \cite{mu2023developer}: they first invited five domain experts to manually label the intents of 8K code snippets and then fine-tuned the CodeBERT model \cite{feng2020codebert} on the labelled data, which was served as a classifier. 
Results show that the fine-tuned model can achieve an F1-score of around 90\%, which is a relatively high value. 
Finally, the authors applied the fine-tuned model to predict the intent category of each comment in the datasets and used the prediction results as the ground-truth labels.
Since manual labelling of such large-scale datasets would be infeasible, we reuse their provided results in our study. 
Also, the training/validation/test partition of the datasets is fixed and the statistics of these two datasets are shown in Table~\ref{tab:dataset}. 
Note that in the table, we do not show the statistics of the validation sets of the two datasets.
This is because our approach does not need to train a model. In contrast, we only retrieve code examples from the training sets (by following Mu \etal \cite{mu2023developer}) with or without customized strategies and evaluate the effectiveness on the test sets.
Therefore, the validation sets are not used in this study. 
Following existing studies \cite{chen2021my,mu2023developer}, we also exclude comments from the {\em others} intent category in our evaluation because these comments are considered as unspecified or ambiguous.


\input{tables/dataset_statistics}

\subsection{Evaluation Metrics}

To evaluate the performance of the Codex model on code summarization, we exploit the common metrics including BLEU \cite{papineni2002bleu}, ROUGE-L \cite{lin2004rouge} and METEOR \cite{banerjee2005meteor}. 
BLEU (Bilingual Evaluation Understudy) \cite{papineni2002bleu} is a commonly-used evaluation metric in the code comment generation studies \cite{iyer2016summarizing,hu2018deep,wan2018improving,mu2023developer}, which measures the similarity between one sentence to a set of reference sentences using constituent n-grams precision scores. 
ROUGE denotes the Recall-oriented Understudy for Gisting Evaluation \cite{lin2004rouge}. It computes the count of several overlapping units such as n-grams, word pairs, and sequences. ROUGE has several different variants from which we consider the most popular one ROUGE-L \cite{lin2021improving,bansal2021project,mu2023developer}, which is calculated based on the longest common subsequence (LCS).
METEOR \cite{banerjee2005meteor}, which denotes the Metric for Evaluation of Translation with Explicit ORdering, is another widely used metric to evaluate the quality of generated code summaries \cite{hu2020deep,wang2020reinforcement,mu2023developer}. METEOR evaluates the generated summary by aligning it to the reference summary and calculating the similarity scores based on the unigram matching.

\subsection{Experiment Settings}

In our experiments, beyond the zero-shot and one-shot settings, we choose to use five and ten code demonstrations for the few-shot setting. We cannot use too many code demonstrations since the input length is restricted by the context window limit. Therefore, we decide to provide the model with ten examples at most.
The baseline for comparison is DOME \cite{mu2023developer} since it is so far the only approach that can address the multi-intent comment generation task.
For running our experiments, we use the latest Codex model {\mycode code-davinci-002}. \footnote{https://platform.openai.com/docs/models/codex} We set the temperature as the default value, 0.5, to get a well-defined answer from Codex. We run all the experiments on an Hygon C86 7385 32-core CPU 2.50GHz machine with 2TB RAM. 
The running OS platform is Ubuntu 18.04.


It is important to note that both the results of RQ1 and RQ2 are subject to randomness. RQ2 is affected by the sampling process, while RQ1 is further influenced by the selection of demonstrations. To address this issue, we repeated each setting one hundred times and reported the average values in the paper. Therefore, the results of RQ1 and RQ2 can be regarded as the expected average effectiveness of Codex under specific settings.
In contrast, RQ3 investigates whether better results can be achieved by leveraging the diversity of sampling results. To accomplish this, we repeated the experiments one hundred times and applied our reranking strategy based on the obtained results. The results of this RQ can thus be considered as the optimal achievable effectiveness of Codex.






%% file: tables/dataset_statistics.tex

\begin{table}[!t]
\caption{The statistics of our evaluation datasets.}
\begin{tabular}{c|cc|cc}
\hline
\multirow{2}{*}{Dataset} & \multicolumn{2}{c|}{Funcom}                                & \multicolumn{2}{c}{TLC}                                \\ \cline{2-5} 
                         & \multicolumn{1}{c|}{Train}     & \multicolumn{1}{c|}{Test} & \multicolumn{1}{c|}{Train}  & \multicolumn{1}{c}{Test} \\ \hline
What                     & \multicolumn{1}{c|}{685,992}   & 44,330                    & \multicolumn{1}{c|}{28,991} & 2,724                     \\ \hline
Why                      & \multicolumn{1}{c|}{152,026}   & 8,402                     & \multicolumn{1}{c|}{5,935}  & 381                       \\ \hline
How-to-use               & \multicolumn{1}{c|}{24,648}    & 1,233                     & \multicolumn{1}{c|}{838}    & 48                        \\ \hline
How-it-is-done           & \multicolumn{1}{c|}{146,571}   & 6,466                     & \multicolumn{1}{c|}{11,478} & 687                       \\ \hline
Property                 & \multicolumn{1}{c|}{166,459}   & 8,326                     & \multicolumn{1}{c|}{5,016}  & 396                       \\ \hline
Total                    & \multicolumn{1}{c|}{1,175,696} & 68,757                    & \multicolumn{1}{c|}{52,258} & 4,236                     \\ \hline
\end{tabular}
\label{tab:dataset}
\end{table}

%% file: 4.evaluation.tex
\section{Study Results}
\label{sec:eval}


\input{tables/RQ1_results}

\subsection{RQ1: the Effectiveness of Vanilla In-Context Learning}

Table~\ref{tab:rq1_results} lists the results of DOME and Codex on the multi-intent comment generation task.
For Codex, the results of using 0, 1, 5, and 10 demonstration examples are respectively illustrated. 
Generally, we observe that the effectiveness of in-context learning will be better with the number of code demonstrations increases. For instance, for the ``what'' intent, the BLEU value of Codex is 19.3\% when no code demonstration is used while this values increases to 34.5\% when using ten examples, on the Funcom dataset.
This is within our expectation because more examples will provide more guidance for the model about the on-going task. 
When compared with the state-of-the-art DOME, we note that the effectiveness of \textbf{zero-shot and one-shot learning} is lower than that of DOME. For instance, the average BLEU values of zero-shot learning on the two datasets are 21.2\% and 18.8\%, respectively, while the corresponding values of DOME are 31.8\% and 22.2\%.  
This indicates that without enough code demonstrations, the potential of LLMs on the multi-intent comment generation task may not be fully leveraged. 

\find{
{\bf Finding-1.} 
Zero-shot and one-shot learning may not fully exploit the potential of the LLMs and their effectiveness is sub-optimal compared with that of the DOME approach.
}

When the number of code demonstrations comes up to five, we observe the effectiveness of Codex is competitive to DOME: the values with respect to the ROUGE-L and METEOR metrics are higher than those of DOME while the BLEU values are sightly lower.
A potential reason is that the BLEU metric excessively focuses on measuring n-gram overlapping.
In concrete, it requires strict consistency (\ie the n-grams must be identical), which is difficult for models that have not been fine-tuned to achieve perfect alignment with the references. 
In contrast, the ROUGE-L and METEOR metrics release this requirement by focusing on the longest common subsequence and considering other features such as the word order in addition to n-grams, respectively.
Nonetheless, when the number of code demonstrations reaches ten, Codex outperforms DOME consistently with respect to all the three metrics and two datasets.
Specifically, the average values of Codex with respect to the three metrics are 33.4\%/76.1\%/24.1\% and 27.2\%/66.7\%/19.2\% on the Funcom and TLC datasets, respectively.
Such performances outperform the state-of-the-art DOME by 5.0\%/79.1\%/17.6\% and 22.5\%/81.8\%/16.4\%, respectively, on the two datasets.
We also find that the performance of different approaches varies across the intent categories: generally, all the approaches have relatively low performances on the ``how-it-is-done'' category.
Such a finding is consistent with the results from the existing study \cite{chen2021my}.

\find{
{\bf Finding-2.}
When the LLM is adequately prompted, its performance will exceed that of the state-of-the-art supervised learning approach.
For instance, when the number of demonstrations is ten, the average ROUGE-L values of Codex on the two datasets are 76.1\%/66.7\%, respectively, outperforming DOME by 79.1\%/81.8\%.
}

\input{tables/RQ2_results}

\subsection{RQ2: the Effectiveness of Demonstration Selection}

The results of different retrieval-based demonstration selection strategies are shown in Table \ref{tab:rq2_results}. The zero-shot setting is excluded from this table since it does not use any code demonstration.
We observe that the demonstration selections based on both token and semantic similarities significantly improve the performances compared with the vanilla random selection.
For instance, when the number of selected examples is ten, the BLEU values of Codex on the Funcom and TLC datasets are 33.4\% and 27.2\%, respectively; while such values increase to 64.5\% (65.9\%) and 60.7\% (62.8\%) when the examples are selected based on token (semantic) similarities, with the relative improvements being 93\% (97\%) and 123\% (131\%).
We also note that such performance improvements are universal (\ie can be observed on each dataset no matter how many code examples are used).
Moreover, we note that if similar examples are provided, the performance of 1-shot learning is even better than that of the vanilla 10-shot learning (\eg the BLEU values on the Funcom dataset are 39.2\% and 33.4\%, respectively).
Such results indicate the importance of the demonstration quality in the in-context learning: the model's performance could be improved if the given prompt is similar to the on-going task.

{\bf Case analysis.}
For qualitative analysis, we present one case to show how the similar code helps to rectify the generated comment of Codex, which is shown in Figure \ref{fig:RQ2example}.
Given the test code whose oracle comment is ``Plays previous video in playlist'', Codex with random selection generates a semantically-irrelevant comment ``Plays the next song or video''.
This comment is inappropriate since the attributive ``next'' is wrong (the oracle is ``previous'') and will thus mislead the potential maintainer of the code.
Fortunately, after using the semantic-based demonstration selection strategy, Codex generates a comment that is semantically-identical to the oracle, \ie ``Plays the previous video in your playlist''. The achieved BLEU score reaches 73.1\%, which is a relatively high performance. 
By investigating the most semantically-similar code in the corpus (listed in the bottom of the figure), we find that one potential reason for the success of Codex is that the example code shows it the attributive could come from the method name.
Specifically, the comment for the semantically-similar code is ``Play the first item'' and ``first'' is a token from the method name.
With this example in mind, Codex generates the correct attributive ``previous'', which can also be extracted from the method name.

\find{
{\bf Finding-3.}
Both token-based and semantic-based demonstration selection strategies can improve the effectiveness of Codex to a large extent.
}


\begin{figure}[!t]
\begin{lstlisting}[mathescape,label={fig:RQ2example}]
// Test Code:
private void playPrevious() {
		if (mediaType == ItemType.YOUTUBE_MEDIA_TYPE_VIDEO) {
			restartVideo();
			return;
		}
		if (currentSongIndex - 1 >= 0) {
			currentSongIndex--;
		} else { 
			currentSongIndex = youTubeVideos.size() - 1;
		}
		videoItem = youTubeVideos.get(youTubeVideos.size() - 1);
		playVideo();
	}
// Ground-truth Comment: Plays previous video in playlist.
// Codex-1-shot: Plays the next song or video.
// Codex-1-shot (Selection): Plays the previous video in your playlist.
----------------------------------------------------------------------------
// Top-1 Semantic-Similar Code
// Comment: Play the first item in the audio queue.
private void playFirstInQueue() {
        AudioQueueItem queueItem = mAudioQueue.poll();
        try {
            mMediaPlayer.setDataSource(this, queueItem.mUri);
        } catch (IOException e) {
            e.printStackTrace();
            endPlayback();
            return;
        }
        mMediaPlayer.setOnCompletionListener(queueItem.mListener);
        try {
            mMediaPlayer.prepare();
        } catch (IOException e) {
            e.printStackTrace();
            endPlayback();
            return;
        }
        mMediaPlayer.start();
    }

\end{lstlisting}

\captionof{figure}{An illustrative example to show how semantic-based selection helps improve the comment generation compared with the random selection.}\label{fig:RQ2example}
\end{figure}

When it comes to the comparison between the two selection strategies, we find that no strategy can consistently outperform the other under all the settings.
For instance, when using one-shot learning, the performance of the token-based selection is better than that of the semantic-based selection on average; and vice versa when using few-shot learning (\ie the number of examples are five or ten).
Moreover, even if the semantic-based selection generally has a better performance when the number of examples is ten, it can also be outperformed by the token-based one under certain settings. For instance, on the {\em what} intent, the BLEU values of the token-based selection are 50.5\% and 44.8\%, respectively, on the two datasets, exceeding those of the semantic-based selection, which are 40.4\% and 40.2\%.

\find{
{\bf Finding-4.}
No demonstration selection strategy can consistently outperform its alternative. The effectiveness depends on the detailed settings (\eg the number of examples and the intents).
}

\input{tables/RQ3_results}

\subsection{RQ3: the Effectiveness of Reranking}

The results of different reranking strategies are shown in Table~\ref{tab:rq3_results}.
Due to the space limitation, we list the results of 1-shot and 10-shot learning. For 1-shot, we also combine different reranking strategies with token-based demonstration selection since according to the results from the above section, this selection strategy achieves better results on 1-shot.
Similarly, for 10-shot, we combine different reranking strategies with semantic-based demonstration selection.
Results show that both reranking strategies help boost the performance of Codex slightly.
For instance, for 1-shot learning, the token-based reranking strategy increases the BLEU values on the Funcom and TLC datasets from 23.1\% and 21.1\% to 29.1\% and 26.8\%, while the semantic-based strategy further achieves 30.2\% and 27.2\% on the two datasets.
We also note that the reranking can enhance the results no matter whether the demonstration selection is used. 
The best-performing model variant, \ie the 10-shot learning with semantic-based demonstration selection and token-based reranking, achieves BLEU scores of 72.4\% and 68.8\% on the two datasets on average, outperforming the state-of-the-art DOME by 128\% and 210\%, respectively (cf. Table~\ref{tab:rq1_results}).

\begin{figure}[!t]
\begin{lstlisting}[mathescape,label={fig:RQ3example}]
// Test Code:
public boolean isDurableSubscriber() {
  return !StringsUtils.isEmpty(m_durableSubscriptionName);
}
// Ground-truth Comment: Determines if the subscriber is durable.
----------------------------------------------------------------------------
/*
Codex-1-shot Top-5 Generated Comments: 
1.Returns true if this subscription has the subscribtion type DURABLE_EXPLICIT or DURABLE_IMPLICIT.
2.Indicates whether or not the Endpoint is a durable subscriber.
3.Returns TRUE if this is a durable subscription and FALSE otherwise.
4.Determines whether or not the subscriber is durable.
5.Can a durable customer install the said customer.
*/
----------------------------------------------------------------------------
// Top-1 Lexical-Similar Code
// Comment: Determines whether or not this subscription is to all stream or to a specific stream.
public boolean isSubscribedToAll() {
    return isNullOrEmpty(streamId);
}
----------------------------------------------------------------------------
// Top-1 result after token-based rerank: 
// Determines whether or not the subscriber is durable.
\end{lstlisting}

\captionof{figure}{An illustrative example to show how our re-ranking strategy helps improve the comment generation.}\label{fig:RQ3example}
\end{figure}

{\bf Case analysis.}
We present another case to show how the reranking strategy helps select more qualified comments,
which is shown in Figure \ref{fig:RQ3example}.
In this figure, we demonstrate the top-5 generated comments from Codex.
The first generated comment is semantically vague since it fails to explicitly explain the meaning of the words {\mycode DURABLE\_EXPLICIT} and {\mycode DURABLE\_IMPLICIT}.
Similarly, the second generated comment may also mislead developers since it is unclear what is an {\mycode Endpoint}, which does not occur in the source code.
The third and forth generated comments share a similar meaning but are expressed in different ways, and they are both semantically-identical to the oracle comment.
After using the token-based similar code selection, a code snippet with the comment ``Determines whether or not ...'' is utilized to help rerank the original results.
Due to a large degree of token overlap with the reference comment, the forth generated comment from Codex is used as the final result according to the token-based reranking strategy.
Compared with the original top-1 result, the BLEU value is increased from 23.4\% to 68.6\%.

\find{
{\bf Finding-5.} 
Both token-based and semantic-based reranking strategies can further enhance the performance of Codex. 
}

As for the comparison between the two reranking strategies, we again observe that no one can consistently outperform its alternative.
Generally, token-based reranking works better when combined with demonstration selections while semantic-based reranking works better when no demonstration selection is adopted.
There are, however, some corner cases. For instance, for the ``what'' intent category, semantic-based reranking performs better when combined with demonstration selections.

\find{
{\bf Finding-6.}
No reranking strategy can consistently outperform its alternative.
}

%% file: tables/RQ1_results.tex
\begin{table}[!t]
\caption{The results of Codex on multi-intent comment generation using zero-shot, one-shot, and few-shot learning (in \%).}
\resizebox{0.99\linewidth}{!}{
\begin{tabular}{c|l|lll|lll}
\hline
\multicolumn{1}{c|}{\multirow{2}{*}{Intent}} & \multicolumn{1}{c|}{\multirow{2}{*}{Method}}                                  & \multicolumn{3}{c|}{Funcom}                                       & \multicolumn{3}{c}{TLC}                                          \\ \cline{3-8} 
\multicolumn{1}{l|}{}                        & \multicolumn{1}{c|}{}                                                         & \multicolumn{1}{l}{BLEU} & \multicolumn{1}{l}{ROUGE-L} & METEOR & \multicolumn{1}{l}{BLEU} & \multicolumn{1}{l}{ROUGE-L} & METEOR \\ \hline
\multirow{5}{*}{What}                         & DOME                                                                          & \multicolumn{1}{l}{33.3} & \multicolumn{1}{c}{41.7}    & 20.5   & \multicolumn{1}{l}{25.4} & \multicolumn{1}{c}{39.6}    & 18.2   \\ \cline{2-8} 
& Codex-0-shot                                                               & \multicolumn{1}{l}{19.3} & \multicolumn{1}{c}{23.5}    & 10.8   & \multicolumn{1}{l}{17.8}     & \multicolumn{1}{c}{16.4}        &  15.5      \\ \cline{2-8} 
& Codex-1-shot                                                               & \multicolumn{1}{l}{23.8} & \multicolumn{1}{c}{27.6}    & 21.5   & \multicolumn{1}{l}{22.5}     & \multicolumn{1}{c}{20.6}        &  17.4      \\ \cline{2-8} 
& Codex-5-shot                                                               & \multicolumn{1}{l}{27.3} & \multicolumn{1}{c}{41.8}    & 24.9   & \multicolumn{1}{l}{25.7}     & \multicolumn{1}{c}{37.4}        &  19.9      \\ \cline{2-8} 
& Codex-10-shot                                                              & \multicolumn{1}{l}{\textbf{34.5}} & \multicolumn{1}{c}{\textbf{58.6}}    & \textbf{26.8}   & \multicolumn{1}{l}{\textbf{32.4}}     & \multicolumn{1}{c}{\textbf{45.6}}        &  \textbf{23.1}      
       \\ \hline
\multirow{5}{*}{Why}                         & DOME                                                                          & \multicolumn{1}{l}{33.0} & \multicolumn{1}{c}{42.3}    & 20.5   & \multicolumn{1}{l}{21.9} & \multicolumn{1}{c}{35.3}    & 15.7   \\ \cline{2-8} 
& Codex-0-shot                                                               & \multicolumn{1}{l}{21.7} & \multicolumn{1}{c}{20.3}    & 11.4   & \multicolumn{1}{l}{19.6}     & \multicolumn{1}{c}{17.8}        & 9.6       \\ \cline{2-8} 
& Codex-1-shot                                                               & \multicolumn{1}{l}{22.9} & \multicolumn{1}{c}{28.8}    & 12.9   & \multicolumn{1}{l}{20.8}     & \multicolumn{1}{c}{23.2}        & 11.9       \\ \cline{2-8} 
& Codex-5-shot                                                               & \multicolumn{1}{l}{27.5} & \multicolumn{1}{c}{45.8}    & 16.9   & \multicolumn{1}{l}{24.1}     & \multicolumn{1}{c}{40.6}        &    13.5    \\ \cline{2-8} 
& Codex-10-shot                                                              & \multicolumn{1}{l}{\textbf{34.8}} & \multicolumn{1}{c}{\textbf{76.1}}    & \textbf{22.6}   & \multicolumn{1}{l}{\textbf{26.2}}     & \multicolumn{1}{c}{\textbf{64.6}}        &   \textbf{15.8}      \\ \hline

\multirow{5}{*}{How-to-use}                         & DOME                                                                          & \multicolumn{1}{l}{31.6} & \multicolumn{1}{c}{39.3}    & 19.3   & \multicolumn{1}{l}{17.1} & \multicolumn{1}{c}{26.1}    & 12.3   \\ \cline{2-8} 
& Codex-0-shot                                                               & \multicolumn{1}{l}{22.3} & \multicolumn{1}{c}{11.1}    & 16.8   & \multicolumn{1}{l}{21.2}     & \multicolumn{1}{c}{10.9}        & 12.2       \\ \cline{2-8} 
& Codex-1-shot                                                               & \multicolumn{1}{l}{23.1} & \multicolumn{1}{c}{18.9}    & 17.5   & \multicolumn{1}{l}{21.8}     & \multicolumn{1}{c}{16.6}        &  14.4      \\ \cline{2-8} 
& Codex-5-shot                                                               & \multicolumn{1}{l}{27.9} & \multicolumn{1}{c}{48.6}    & 19.8    & \multicolumn{1}{l}{24.4}     & \multicolumn{1}{c}{40.5}        &  15.7      \\ \cline{2-8} 
& Codex-10-shot                                                              & \multicolumn{1}{l}{\textbf{33.3}} & \multicolumn{1}{c}{\textbf{84.6}}    & \textbf{22.3}   & \multicolumn{1}{l}{\textbf{26.9}}     & \multicolumn{1}{c}{\textbf{76.4}}        &   \textbf{17.3}     \\ \hline

\multirow{5}{*}{How-it-is-done}                         & DOME                                                                          & \multicolumn{1}{l}{26.9} & \multicolumn{1}{c}{39.5}    & 17.6   & \multicolumn{1}{l}{20.4} & \multicolumn{1}{c}{36.6}    & 14.7   \\ \cline{2-8} 
& Codex-0-shot                                                               & \multicolumn{1}{l}{18.9} & \multicolumn{1}{c}{37.9}    & 9.8   & \multicolumn{1}{l}{16.8}     & \multicolumn{1}{c}{32.1}        & 9.6       \\ \cline{2-8} 
& Codex-1-shot                                                               & \multicolumn{1}{l}{21.0} & \multicolumn{1}{c}{39.6}    & 13.5   & \multicolumn{1}{l}{19.1}     & \multicolumn{1}{c}{36.4}        &   12.1     \\ \cline{2-8} 
& Codex-5-shot                                                               & \multicolumn{1}{l}{24.8} & \multicolumn{1}{c}{49.2}    & 16.2   & \multicolumn{1}{l}{21.1}     & \multicolumn{1}{c}{52.7}        &   12.8     \\ \cline{2-8} 
& Codex-10-shot                                                              & \multicolumn{1}{l}{\textbf{28.4}} & \multicolumn{1}{c}{\textbf{79.3}}    & \textbf{19.5}   & \multicolumn{1}{l}{\textbf{21.9}}     & \multicolumn{1}{c}{\textbf{66.7}}        & \textbf{14.9}        \\ \hline

\multirow{5}{*}{Property}                         & DOME                                                                          & \multicolumn{1}{l}{34.1} & \multicolumn{1}{c}{49.4}    & 24.3   & \multicolumn{1}{l}{26.0} & \multicolumn{1}{c}{45.7}    & 21.2   \\ \cline{2-8} 
& Codex-0-shot                                                               & \multicolumn{1}{l}{23.7} & \multicolumn{1}{c}{33.3}    &  13.2  & \multicolumn{1}{l}{18.8}     & \multicolumn{1}{c}{28.8}        & 9.5       \\ \cline{2-8} 
& Codex-1-shot                                                               & \multicolumn{1}{l}{24.7} & \multicolumn{1}{c}{38.4}    &  15.8  & \multicolumn{1}{l}{21.3}     & \multicolumn{1}{c}{33.6}        & 12.4       \\ \cline{2-8} 
& Codex-5-shot                                                               & \multicolumn{1}{l}{29.7} & \multicolumn{1}{c}{79.2}    & 25.2   & \multicolumn{1}{l}{26.5}     & \multicolumn{1}{c}{78.4}        & 22.3       \\ \cline{2-8} 
& Codex-10-shot                                                              & \multicolumn{1}{l}{\textbf{36.2}} & \multicolumn{1}{c}{\textbf{81.9}}    &  \textbf{29.4}  & \multicolumn{1}{l}{\textbf{28.7}}     & \multicolumn{1}{c}{\textbf{80.3}}        &   \textbf{24.7}         \\ \hline
\multirow{5}{*}{Average}                         & DOME                                                                          & \multicolumn{1}{l}{31.8} & \multicolumn{1}{c}{42.5}    & 20.5   & \multicolumn{1}{l}{22.2} & \multicolumn{1}{c}{36.7}    & 16.5   \\ \cline{2-8} 
& Codex-0-shot                                                               & \multicolumn{1}{l}{21.2} & \multicolumn{1}{c}{25.2}    & 12.4   & \multicolumn{1}{l}{18.8}     & \multicolumn{1}{c}{21.2}        &  11.3      \\ \cline{2-8} 
& Codex-1-shot                                                               & \multicolumn{1}{l}{23.1} & \multicolumn{1}{c}{30.7}    & 16.2   & \multicolumn{1}{l}{21.1}     & \multicolumn{1}{c}{26.1}        & 13.6       \\ \cline{2-8} 
& Codex-5-shot                                                               & \multicolumn{1}{l}{27.4} & \multicolumn{1}{c}{52.9}    &  20.6  & \multicolumn{1}{l}{24.4}     & \multicolumn{1}{c}{49.9}        & 16.8       \\ \cline{2-8} 
& Codex-10-shot                                                              & \multicolumn{1}{l}{\textbf{33.4}} & \multicolumn{1}{c}{\textbf{76.1}}    & \textbf{24.1}   & \multicolumn{1}{l}{\textbf{27.2}}     & \multicolumn{1}{c}{\textbf{66.7}}        &   \textbf{19.2}    
       \\ \hline
\end{tabular}
}
\label{tab:rq1_results}
\end{table}

%% file: tables/RQ2_results.tex
\begin{table*}[!t]
\caption{The results of different retrieval-based demonstration selection strategies (in \%).}
\resizebox{0.6\linewidth}{!}{
\begin{tabular}{c|l|lll|lll}
\hline
\multicolumn{1}{c|}{\multirow{2}{*}{Intent}} & \multicolumn{1}{c|}{\multirow{2}{*}{Method}}                                  & \multicolumn{3}{c|}{Funcom}                                       & \multicolumn{3}{c}{TLC}                                          \\ \cline{3-8} 
\multicolumn{1}{l|}{}                        & \multicolumn{1}{c|}{}                                                         & \multicolumn{1}{l}{BLEU} & \multicolumn{1}{l}{ROUGE-L} & METEOR & \multicolumn{1}{l}{BLEU} & \multicolumn{1}{l}{ROUGE-L} & METEOR \\ \hline
\multirow{9}{*}{What}           & \begin{tabular}[c]{@{}l@{}}Codex-1-shot \end{tabular}               & \multicolumn{1}{l}{23.8} & \multicolumn{1}{c}{27.6}    &  21.5  & \multicolumn{1}{l}{22.5}     & \multicolumn{1}{c}{20.6}        & 17.4       \\     & \begin{tabular}[c]{@{}l@{}}Codex-1-shot ($Selection_{token}$)\end{tabular}               & \multicolumn{1}{l}{\textbf{39.5}} & \multicolumn{1}{c}{\textbf{84.6}}    & 35.0   & \multicolumn{1}{l}{\textbf{35.6}}     & \multicolumn{1}{c}{\textbf{79.9}}        & 31.4       \\  
& \begin{tabular}[c]{@{}l@{}}Codex-1-shot ($Selection_{semantic}$)\end{tabular}               & \multicolumn{1}{l}{36.7} & \multicolumn{1}{c}{74.5}    & \textbf{36.1}   & \multicolumn{1}{l}{33.9}     & \multicolumn{1}{c}{71.6}        & \textbf{32.8}       \\ \cline{2-8} 
& \begin{tabular}[c]{@{}l@{}}Codex-5-shot \end{tabular}               & \multicolumn{1}{l}{27.3} & \multicolumn{1}{c}{41.8}    & 24.9   & \multicolumn{1}{l}{25.7}     & \multicolumn{1}{c}{37.4}        & 19.9       \\  
& \begin{tabular}[c]{@{}l@{}}Codex-5-shot ($Selection_{token}$)\end{tabular}               & \multicolumn{1}{l}{41.0} & \multicolumn{1}{c}{82.3}    & \textbf{41.3}   & \multicolumn{1}{l}{38.6}     & \multicolumn{1}{c}{76.8}        & 37.7       \\ 
& \begin{tabular}[c]{@{}l@{}}Codex-5-shot ($Selection_{semantic}$)\end{tabular} & \multicolumn{1}{l}{\textbf{41.1}} & \multicolumn{1}{c}{\textbf{82.9}}    & 39.3   & \multicolumn{1}{l}{\textbf{39.1}}     & \multicolumn{1}{c}{\textbf{78.9}}        & \textbf{38.3}   \\ \cline{2-8} & 
\begin{tabular}[c]{@{}l@{}}Codex-10-shot\end{tabular}               & \multicolumn{1}{l}{34.5} & \multicolumn{1}{c}{58.6}    & 26.8   & \multicolumn{1}{l}{32.4}     & \multicolumn{1}{c}{45.6}        & 23.1       \\     & \begin{tabular}[c]{@{}l@{}}Codex-10-shot ($Selection_{token}$)\end{tabular}               & \multicolumn{1}{l}{\textbf{50.5}} & \multicolumn{1}{c}{\textbf{90.0}}    & \textbf{48.4}   & \multicolumn{1}{l}{\textbf{44.8}}     & \multicolumn{1}{c}{\textbf{82.6}}        &  \textbf{43.9}      \\  
& \begin{tabular}[c]{@{}l@{}}Codex-10-shot ($Selection_{semantic}$)\end{tabular}               & \multicolumn{1}{l}{40.4} & \multicolumn{1}{c}{84.1}    & 38.7   & \multicolumn{1}{l}{40.2}     & \multicolumn{1}{c}{79.5}& 38.2   
\\ \hline
\multirow{9}{*}{Why}   &                 \begin{tabular}[c]{@{}l@{}}Codex-1-shot\end{tabular}               & \multicolumn{1}{l}{22.9} & \multicolumn{1}{c}{28.8}    & 12.9   & \multicolumn{1}{l}{20.8}     & \multicolumn{1}{c}{23.2}        & 11.9       \\     & \begin{tabular}[c]{@{}l@{}}Codex-1-shot ($Selection_{token}$)\end{tabular}               & \multicolumn{1}{l}{32.8} & \multicolumn{1}{c}{\textbf{72.8}}    & 27.7   & \multicolumn{1}{l}{30.7}     & \multicolumn{1}{c}{\textbf{68.4}}        &   25.5     \\  
& \begin{tabular}[c]{@{}l@{}}Codex-1-shot ($Selection_{semantic}$)\end{tabular}               & \multicolumn{1}{l}{\textbf{33.2}} & \multicolumn{1}{c}{70.9}    & \textbf{28.0}   & \multicolumn{1}{l}{\textbf{31.6}}     & \multicolumn{1}{c}{66.8}        &   \textbf{26.2}     \\  \cline{2-8}
& \begin{tabular}[c]{@{}l@{}}Codex-5-shot \end{tabular}               & \multicolumn{1}{l}{24.2} & \multicolumn{1}{c}{45.5}    & 14.7   & \multicolumn{1}{l}{24.1}     & \multicolumn{1}{c}{40.6}        & 13.5       \\     
& \begin{tabular}[c]{@{}l@{}}Codex-5-shot ($Selection_{token}$)\end{tabular}               & \multicolumn{1}{l}{\textbf{37.8}} & \multicolumn{1}{c}{\textbf{85.0}}    & \textbf{32.9}   & \multicolumn{1}{l}{34.5}     & \multicolumn{1}{c}{78.7}        & 29.8       \\  
& \begin{tabular}[c]{@{}l@{}}Codex-5-shot ($Selection_{semantic}$)\end{tabular} & \multicolumn{1}{l}{37.7} & \multicolumn{1}{c}{82.1}    & 32.5   & \multicolumn{1}{l}{\textbf{35.1}}     & \multicolumn{1}{c}{\textbf{79.3}}        &   \textbf{30.2} \\ \cline{2-8} &
\begin{tabular}[c]{@{}l@{}}Codex-10-shot\end{tabular}               & \multicolumn{1}{l}{34.8} & \multicolumn{1}{c}{76.1}    & 22.6   & \multicolumn{1}{l}{26.2}     & \multicolumn{1}{c}{64.6}        & 15.8       \\     & \begin{tabular}[c]{@{}l@{}}Codex-10-shot ($Selection_{token}$)\end{tabular}               & \multicolumn{1}{l}{74.9} & \multicolumn{1}{c}{\textbf{90.0}}    & \textbf{75.1}   & \multicolumn{1}{l}{72.1}     & \multicolumn{1}{c}{81.4}        &  68.9      \\  
& \begin{tabular}[c]{@{}l@{}}Codex-10-shot ($Selection_{semantic}$)\end{tabular}               & \multicolumn{1}{l}{\textbf{75.0}} & \multicolumn{1}{c}{89.4}    & 74.7  & \multicolumn{1}{l}{\textbf{72.4}}     & \multicolumn{1}{c}{\textbf{81.9}} & \textbf{73.0}
\\ \hline

\multirow{9}{*}{How-to-use}   &
\begin{tabular}[c]{@{}l@{}}Codex-1-shot\end{tabular}               & \multicolumn{1}{l}{23.1} & \multicolumn{1}{c}{18.9}    & 17.5   & \multicolumn{1}{l}{21.8}     & \multicolumn{1}{c}{16.6}        & 14.4       \\     & \begin{tabular}[c]{@{}l@{}}Codex-1-shot ($Selection_{token}$)\end{tabular}               & \multicolumn{1}{l}{\textbf{56.3}} & \multicolumn{1}{c}{\textbf{88.3}}    & \textbf{53.7}   & \multicolumn{1}{l}{\textbf{52.2}}     & \multicolumn{1}{c}{\textbf{81.6}}        &   \textbf{42.8}     \\  
& \begin{tabular}[c]{@{}l@{}}Codex-1-shot ($Selection_{semantic}$)\end{tabular}               & \multicolumn{1}{l}{52.4} & \multicolumn{1}{c}{74.4}    & 47.1   & \multicolumn{1}{l}{46.8}     & \multicolumn{1}{c}{71.5}        &   42.3     \\ \cline{2-8} 
& \begin{tabular}[c]{@{}l@{}}Codex-5-shot \end{tabular}               & \multicolumn{1}{l}{24.2} & \multicolumn{1}{c}{48.1}    & 18.9   & \multicolumn{1}{l}{24.4}     & \multicolumn{1}{c}{40.5}        & 15.7       \\             
& \begin{tabular}[c]{@{}l@{}}Codex-5-shot ($Selection_{token}$)\end{tabular}               & \multicolumn{1}{l}{48.0} & \multicolumn{1}{c}{\textbf{86.4}}    & 45.9   & \multicolumn{1}{l}{43.6}     & \multicolumn{1}{c}{80.3}        & 37.2       \\  
& \begin{tabular}[c]{@{}l@{}}Codex-5-shot ($Selection_{semantic}$)\end{tabular} & \multicolumn{1}{l}{\textbf{68.7}} & \multicolumn{1}{c}{86.2}    & \textbf{63.6}   & \multicolumn{1}{l}{\textbf{66.4}}     & \multicolumn{1}{c}{\textbf{84.5}}        & \textbf{58.4}       \\  \cline{2-8} &

\begin{tabular}[c]{@{}l@{}}Codex-10-shot\end{tabular}               & \multicolumn{1}{l}{33.3} & \multicolumn{1}{c}{84.6}    & 22.3   & \multicolumn{1}{l}{26.9}     & \multicolumn{1}{c}{76.4}        & 17.3       \\     & \begin{tabular}[c]{@{}l@{}}Codex-10-shot ($Selection_{token}$)\end{tabular}               & \multicolumn{1}{l}{69.6} & \multicolumn{1}{c}{91.2}    & 70.7   & \multicolumn{1}{l}{66.4}     & \multicolumn{1}{c}{84.3}        &  68.2      \\  
& \begin{tabular}[c]{@{}l@{}}Codex-10-shot ($Selection_{semantic}$)\end{tabular}               & \multicolumn{1}{l}{\textbf{76.3}} & \multicolumn{1}{c}{\textbf{91.2}}    & \textbf{77.4}   & \multicolumn{1}{l}{\textbf{71.6}}     & \multicolumn{1}{c}{\textbf{85.4}} & \textbf{73.6} 
\\ \hline
\multirow{9}{*}{How-it-is-done}                         &                 \begin{tabular}[c]{@{}l@{}}Codex-1-shot\end{tabular}               & \multicolumn{1}{l}{21.0} & \multicolumn{1}{c}{39.6}    & 13.5   & \multicolumn{1}{l}{19.1}     & \multicolumn{1}{c}{36.4}        & 12.1       \\     & \begin{tabular}[c]{@{}l@{}}Codex-1-shot ($Selection_{token}$)\end{tabular}               & \multicolumn{1}{l}{\textbf{31.9}} & \multicolumn{1}{c}{\textbf{72.9}}    & 25.8   & \multicolumn{1}{l}{\textbf{28.6}}     & \multicolumn{1}{c}{\textbf{69.4}}        & 24.7       \\  
& \begin{tabular}[c]{@{}l@{}}Codex-1-shot ($Selection_{semantic}$)\end{tabular}               & \multicolumn{1}{l}{30.5} & \multicolumn{1}{c}{69.6}    & \textbf{27.6}   & \multicolumn{1}{l}{28.2}     & \multicolumn{1}{c}{68.7}        &  \textbf{25.9}      \\ \cline{2-8} 
& \begin{tabular}[c]{@{}l@{}}Codex-5-shot \end{tabular}               & \multicolumn{1}{l}{22.5} & \multicolumn{1}{c}{48.9}    & 13.7   & \multicolumn{1}{l}{21.1}     & \multicolumn{1}{c}{52.7}        & 12.8       \\ 
& \begin{tabular}[c]{@{}l@{}}Codex-5-shot ($Selection_{token}$)\end{tabular}               & \multicolumn{1}{l}{\textbf{33.7}} & \multicolumn{1}{c}{\textbf{85.7}}    &  \textbf{30.8}  & \multicolumn{1}{l}{\textbf{29.7}}     & \multicolumn{1}{c}{\textbf{78.4}}        & \textbf{26.8}       \\  
& \begin{tabular}[c]{@{}l@{}}Codex-5-shot ($Selection_{semantic}$)\end{tabular} & \multicolumn{1}{l}{32.9} & \multicolumn{1}{c}{80.0}    &  27.5  & \multicolumn{1}{l}{28.3}     & \multicolumn{1}{c}{73.9}        & 25.1 
\\ \cline{2-8} &
\begin{tabular}[c]{@{}l@{}}Codex-10-shot\end{tabular}               & \multicolumn{1}{l}{28.4} & \multicolumn{1}{c}{79.3}    & 19.5   & \multicolumn{1}{l}{21.9}     & \multicolumn{1}{c}{66.7}        & 14.9       \\     & \begin{tabular}[c]{@{}l@{}}Codex-10-shot ($Selection_{token}$)\end{tabular}               & \multicolumn{1}{l}{47.9} & \multicolumn{1}{c}{84.6}    & 49.6   & \multicolumn{1}{l}{45.2}     & \multicolumn{1}{c}{80.8}        & 47.7       \\ 
& \begin{tabular}[c]{@{}l@{}}Codex-10-shot ($Selection_{semantic}$)\end{tabular}               & \multicolumn{1}{l}{\textbf{51.6}} & \multicolumn{1}{c}{\textbf{86.4}}    & \textbf{50.8}   & \multicolumn{1}{l}{\textbf{48.9}}     & \multicolumn{1}{c}{\textbf{82.9}} & \textbf{47.9} 
\\ \hline

\multirow{9}{*}{Property}               &
\begin{tabular}[c]{@{}l@{}}Codex-1-shot\end{tabular}               & \multicolumn{1}{l}{24.7} & \multicolumn{1}{c}{38.4}    & 15.8   & \multicolumn{1}{l}{21.3}     & \multicolumn{1}{c}{33.6}        & 12.4       \\     & \begin{tabular}[c]{@{}l@{}}Codex-1-shot ($Selection_{token}$)\end{tabular}               & \multicolumn{1}{l}{35.7} & \multicolumn{1}{c}{67.7}    & 33.1   & \multicolumn{1}{l}{33.2}     & \multicolumn{1}{c}{\textbf{64.9}}        & 30.8       \\  
& \begin{tabular}[c]{@{}l@{}}Codex-1-shot ($Selection_{semantic}$)\end{tabular}               & \multicolumn{1}{l}{\textbf{36.4}} & \multicolumn{1}{c}{\textbf{80.0}}    & \textbf{35.9}   & \multicolumn{1}{l}{\textbf{34.9}}     & \multicolumn{1}{c}{62.8}        & \textbf{32.4}       \\ \cline{2-8} 
& \begin{tabular}[c]{@{}l@{}}Codex-5-shot \end{tabular}               & \multicolumn{1}{l}{29.7} & \multicolumn{1}{c}{79.2}    & 25.2   & \multicolumn{1}{l}{26.5}     & \multicolumn{1}{c}{78.4}        & 22.3       \\                
& \begin{tabular}[c]{@{}l@{}}Codex-5-shot ($Selection_{token}$)\end{tabular}               & \multicolumn{1}{l}{\textbf{45.1}} & \multicolumn{1}{c}{\textbf{89.2}}    & 43.2   & \multicolumn{1}{l}{\textbf{41.5}}     & \multicolumn{1}{c}{\textbf{85.4}}        & \textbf{40.6}       \\  
& \begin{tabular}[c]{@{}l@{}}Codex-5-shot ($Selection_{semantic}$)\end{tabular} & \multicolumn{1}{l}{43.9} & \multicolumn{1}{c}{82.7}    &  40.3  & \multicolumn{1}{l}{39.6}     &  \multicolumn{1}{c}{82.1}        & 38.1 
\\ \cline{2-8} &
\begin{tabular}[c]{@{}l@{}}Codex-10-shot\end{tabular}               & \multicolumn{1}{l}{36.2} & \multicolumn{1}{c}{81.9}    & 29.4   & \multicolumn{1}{l}{28.7}     & \multicolumn{1}{c}{80.3}        & 24.7       \\     & \begin{tabular}[c]{@{}l@{}}Codex-10-shot ($Selection_{token}$)\end{tabular}               & \multicolumn{1}{l}{79.6} & \multicolumn{1}{c}{84.2}    & 75.7   & \multicolumn{1}{l}{74.8}     & \multicolumn{1}{c}{83.9}        & 68.9       \\ 
& \begin{tabular}[c]{@{}l@{}}Codex-10-shot ($Selection_{semantic}$)\end{tabular}               & \multicolumn{1}{l}{\textbf{86.3}} & \multicolumn{1}{c}{\textbf{95.8}}    & \textbf{87.4}   & \multicolumn{1}{l}{\textbf{81.0}}     & \multicolumn{1}{c}{\textbf{86.4}} & \textbf{80.8}
\\ \hline
\multirow{9}{*}{Average}               &
\begin{tabular}[c]{@{}l@{}}Codex-1-shot\end{tabular}               & \multicolumn{1}{l}{23.1} & \multicolumn{1}{c}{30.7}    & 16.2   & \multicolumn{1}{l}{21.1}     & \multicolumn{1}{c}{26.1}        &    13.6    \\     &\begin{tabular}[c]{@{}l@{}}Codex-1-shot ($Selection_{token}$)\end{tabular}               & \multicolumn{1}{l}{\textbf{39.2}} & \multicolumn{1}{c}{\textbf{77.3}}    &  \textbf{35.1}  & \multicolumn{1}{l}{\textbf{36.1}}     & \multicolumn{1}{c}{\textbf{72.8}}        &  31.0      \\  
& \begin{tabular}[c]{@{}l@{}}Codex-1-shot ($Selection_{semantic}$)\end{tabular}               & \multicolumn{1}{l}{37.8} & \multicolumn{1}{c}{73.9}    &  35.0  & \multicolumn{1}{l}{35.1}     & \multicolumn{1}{c}{68.3}        &  \textbf{31.9}      \\ \cline{2-8} 
& \begin{tabular}[c]{@{}l@{}}Codex-5-shot \end{tabular}               & \multicolumn{1}{l}{27.4} & \multicolumn{1}{c}{52.9}    &  20.6  & \multicolumn{1}{l}{24.4}     & \multicolumn{1}{c}{49.9}        & 16.8       \\                
& \begin{tabular}[c]{@{}l@{}}Codex-5-shot ($Selection_{token}$)\end{tabular}               & \multicolumn{1}{l}{41.1} & \multicolumn{1}{c}{\textbf{85.7}}    &  38.8  & \multicolumn{1}{l}{37.6}     & \multicolumn{1}{c}{\textbf{79.9}}        &  34.4      \\  
& \begin{tabular}[c]{@{}l@{}}Codex-5-shot ($Selection_{semantic}$)\end{tabular} & \multicolumn{1}{l}{\textbf{44.9}} & \multicolumn{1}{c}{82.8}    &  \textbf{40.6}  & \multicolumn{1}{l}{\textbf{41.7}}     &  \multicolumn{1}{c}{79.7}        &  \textbf{38.0}
\\ \cline{2-8} &
\begin{tabular}[c]{@{}l@{}}Codex-10-shot\end{tabular}               & \multicolumn{1}{l}{33.4} & \multicolumn{1}{c}{76.1}    & 24.1   & \multicolumn{1}{l}{27.2}     & \multicolumn{1}{c}{66.7}        & 19.2       \\     & \begin{tabular}[c]{@{}l@{}}Codex-10-shot ($Selection_{token}$)\end{tabular}               & \multicolumn{1}{l}{64.5} & \multicolumn{1}{c}{88.0}    & 63.9   & \multicolumn{1}{l}{60.7}     & \multicolumn{1}{c}{82.6}        &   59.5     \\ 
& \begin{tabular}[c]{@{}l@{}}Codex-10-shot ($Selection_{semantic}$)\end{tabular}               & \multicolumn{1}{l}{\textbf{65.9}} & \multicolumn{1}{c}{\textbf{89.4}}    &  \textbf{65.8}  & \multicolumn{1}{l}{\textbf{62.8}}     & \multicolumn{1}{c}{\textbf{83.2}} & \textbf{62.7}
\\ \hline
\end{tabular}
}
\label{tab:rq2_results}
\end{table*}

%% file: tables/RQ3_results.tex
\begin{table*}[!t]
\caption{The results of different reranking strategies (in \%).}
\resizebox{0.7\linewidth}{!}{
\begin{tabular}{c|l|lll|lll}
\hline
\multicolumn{1}{c|}{\multirow{2}{*}{Intent}} & \multicolumn{1}{c|}{\multirow{2}{*}{Method}}                                  & \multicolumn{3}{c|}{Funcom}                                       & \multicolumn{3}{c}{TLC}                                          \\ \cline{3-8} 
\multicolumn{1}{l|}{}                        & \multicolumn{1}{c|}{}                                                         & \multicolumn{1}{l}{BLEU} & \multicolumn{1}{l}{ROUGE-L} & METEOR & \multicolumn{1}{l}{BLEU} & \multicolumn{1}{l}{ROUGE-L} & METEOR \\ \hline
\multirow{12}{*}{what}

& Codex-1-shot                                                               & \multicolumn{1}{l}{23.8} & \multicolumn{1}{c}{27.6}    & 21.5   & \multicolumn{1}{l}{22.5}     & \multicolumn{1}{c}{20.6}        &  17.4      \\  
& Codex-1-shot ($Rerank_{token}$)
& \multicolumn{1}{l}{\textbf{32.2}} & \multicolumn{1}{c}{76.1}    &  \textbf{33.3}  & \multicolumn{1}{l}{\textbf{28.9}}     & \multicolumn{1}{c}{\textbf{72.7}}        &   \textbf{29.3}     \\  
& Codex-1-shot ($Rerank_{semantic}$)
& \multicolumn{1}{l}{29.7} & \multicolumn{1}{c}{ \textbf{76.5}}    &  26.7   & \multicolumn{1}{l}{27.1}     & \multicolumn{1}{c}{71.9}        & 24.8       \\ \cline{2-8} 

& Codex-1-shot ($Selection_{token}$)                                                               & \multicolumn{1}{l}{39.5} & \multicolumn{1}{c}{84.6}    & 35.0   & \multicolumn{1}{l}{35.6}     & \multicolumn{1}{c}{79.9}        & 31.4       \\  
& Codex-1-shot ($Selection_{token} + Rerank_{token}$)
& \multicolumn{1}{l}{44.4} & \multicolumn{1}{c}{84.9}    &  43.4  & \multicolumn{1}{l}{41.8}     & \multicolumn{1}{c}{\textbf{77.6}}        &   38.5     \\  
& Codex-1-shot ($Selection_{token} + Rerank_{semantic}$)
& \multicolumn{1}{l}{\textbf{45.8}} & \multicolumn{1}{c}{\textbf{85.2}}    & \textbf{44.9}   & \multicolumn{1}{l}{\textbf{42.6}}     & \multicolumn{1}{c}{75.8}        &  \textbf{40.8}      \\ \cline{2-8}

& Codex-10-shot                                                              & \multicolumn{1}{l}{34.5} & \multicolumn{1}{c}{58.6}    & 26.8   & \multicolumn{1}{l}{32.4}     & \multicolumn{1}{c}{45.6}        &  \textbf{23.1}    \\

& Codex-10-shot ($Rerank_{token}$)                                            & \multicolumn{1}{l}{36.9} & \multicolumn{1}{c}{84.5}    &  29.3  & \multicolumn{1}{l}{34.8}     & \multicolumn{1}{c}{76.9}        & 26.6    
\\

& Codex-10-shot ($Rerank_{semantic}$)                                            & \multicolumn{1}{l}{\textbf{39.7}} & \multicolumn{1}{c}{\textbf{85.6}}    &  \textbf{36.5}  & \multicolumn{1}{l}{\textbf{37.1}}     & \multicolumn{1}{c}{\textbf{81.0}}        & \textbf{31.8}    \\ \cline{2-8} 
& Codex-10-shot ($Selection_{semantic}$)                                                               & \multicolumn{1}{l}{40.4} & \multicolumn{1}{c}{84.1}    & 38.7   & \multicolumn{1}{l}{40.2}     & \multicolumn{1}{c}{79.5}        &  38.2      \\  
& Codex-10-shot ($Selection_{semantic} + Rerank_{token}$)
& \multicolumn{1}{l}{58.6} & \multicolumn{1}{c}{87.2}    &  61.3  & \multicolumn{1}{l}{56.3}     & \multicolumn{1}{c}{82.9}        &   58.4     \\  
& Codex-10-shot ($Selection_{semantic} + Rerank_{semantic}$)
& \multicolumn{1}{l}{\textbf{60.2}} & \multicolumn{1}{c}{\textbf{89.4}}    & \textbf{64.1}   & \multicolumn{1}{l}{\textbf{58.3}}     & \multicolumn{1}{c}{\textbf{85.2}}        &  \textbf{60.9}       
  
       \\ \hline
\multirow{12}{*}{why}          
& Codex002-1-shot                                                               & \multicolumn{1}{l}{22.9} & \multicolumn{1}{c}{28.8}    &  12.9  & \multicolumn{1}{l}{20.8}     &  \multicolumn{1}{c}{23.2}        & 11.9       \\  
& Codex-1-shot ($Rerank_{token}$)
& \multicolumn{1}{l}{23.5} & \multicolumn{1}{c}{67.6}    & 17.7   & \multicolumn{1}{l}{22.6}     & \multicolumn{1}{c}{62.7}        &  19.4      \\  
& Codex-1-shot ($Rerank_{semantic}$)
& \multicolumn{1}{l}{\textbf{29.2}} & \multicolumn{1}{c}{\textbf{68.0}}    & \textbf{25.7}   & \multicolumn{1}{l}{\textbf{26.7}}     & \multicolumn{1}{c}{\textbf{63.3}}        &  \textbf{20.1}      \\ \cline{2-8} 

& Codex-1-shot ($Selection_{token}$)                                                               & \multicolumn{1}{l}{32.8} & \multicolumn{1}{c}{72.8}    & 27.7   & \multicolumn{1}{l}{30.7}     & \multicolumn{1}{c}{68.4}        & 25.5       \\  
& Codex-1-shot ($Selection_{token} + Rerank_{token}$)
& \multicolumn{1}{l}{36.4} & \multicolumn{1}{c}{81.0}    &  31.6  & \multicolumn{1}{l}{34.4}     & \multicolumn{1}{c}{77.1}        & 28.9       \\  
& Codex-1-shot ($Selection_{token} + Rerank_{semantic}$)
& \multicolumn{1}{l}{\textbf{38.6}} & \multicolumn{1}{c}{\textbf{83.4}}    & \textbf{35.9}   & \multicolumn{1}{l}{\textbf{36.9}}     & \multicolumn{1}{c}{\textbf{80.2}}        &  \textbf{30.3}      \\ \cline{2-8}

& Codex-10-shot                                                              & \multicolumn{1}{l}{34.8} & \multicolumn{1}{c}{76.1}    & 22.6   & \multicolumn{1}{l}{26.2}     & \multicolumn{1}{c}{64.6}        &  15.8    \\

& Codex-10-shot ($Rerank_{token}$)                                            & \multicolumn{1}{l}{\textbf{36.8}} & \multicolumn{1}{c}{\textbf{91.0}}    & \textbf{24.8}  & \multicolumn{1}{l}{\textbf{31.2}}     & \multicolumn{1}{c}{\textbf{86.1}}        &   \textbf{20.9}  
\\

& Codex-10-shot ($Rerank_{semantic}$)                                            & \multicolumn{1}{l}{35.3} & \multicolumn{1}{c}{90.9}    &  23.2  & \multicolumn{1}{l}{30.4}     & \multicolumn{1}{c}{85.2}        & 20.1    \\ \cline{2-8} 
& Codex-10-shot ($Selection_{semantic}$)                                                               & \multicolumn{1}{l}{75.0} & \multicolumn{1}{c}{89.4}    &  74.7  & \multicolumn{1}{l}{72.4}     & \multicolumn{1}{c}{81.9}        & 73.0       \\  
& Codex-10-shot ($Selection_{semantic} + Rerank_{token}$)
& \multicolumn{1}{l}{\textbf{78.3}} & \multicolumn{1}{c}{\textbf{92.4}}    & \textbf{76.6}   & \multicolumn{1}{l}{\textbf{74.8}}     & \multicolumn{1}{c}{\textbf{88.7}}     &   \textbf{74.1}     \\  
& Codex-10-shot ($Selection_{semantic} + Rerank_{semantic}$)
& \multicolumn{1}{l}{76.2} & \multicolumn{1}{c}{90.6}    & 75.3   & \multicolumn{1}{l}{73.5}     & \multicolumn{1}{c}{86.2}        & 73.6        
  
       \\ \hline
\multirow{12}{*}{How-to-use}          
& Codex-1-shot                                                               & \multicolumn{1}{l}{23.1} & \multicolumn{1}{c}{18.9}    & 17.5   & \multicolumn{1}{l}{21.8}     & \multicolumn{1}{c}{16.6}        & 14.4       \\  
& Codex-1-shot ($Rerank_{token}$)
& \multicolumn{1}{l}{25.1} & \multicolumn{1}{c}{62.0}    &  19.7  & \multicolumn{1}{l}{24.2}     & \multicolumn{1}{c}{58.8}        & 17.6       \\  
& Codex-1-shot ($Rerank_{semantic}$)
& \multicolumn{1}{l}{\textbf{28.5}} & \multicolumn{1}{c}{\textbf{63.6}}    & \textbf{22.9}   & \multicolumn{1}{l}{\textbf{26.1}}     & \multicolumn{1}{c}{\textbf{61.3}}        &   \textbf{18.8}     \\ \cline{2-8} 

& Codex-1-shot ($Selection_{token}$)                                                               & \multicolumn{1}{l}{56.3} & \multicolumn{1}{c}{88.3}    & 53.7   & \multicolumn{1}{l}{52.2}     & \multicolumn{1}{c}{81.6}        & 42.8       \\  
& Codex-1-shot ($Selection_{token} + Rerank_{token}$)
& \multicolumn{1}{l}{\textbf{63.8}} & \multicolumn{1}{c}{\textbf{90.7}}    &  \textbf{66.3}  & \multicolumn{1}{l}{\textbf{60.6}}     & \multicolumn{1}{c}{\textbf{85.3}}        & \textbf{59.7}       \\  
& Codex-1-shot ($Selection_{token} + Rerank_{semantic}$)
& \multicolumn{1}{l}{61.1} & \multicolumn{1}{c}{85.7}    &  60.6  & \multicolumn{1}{l}{58.4}     & \multicolumn{1}{c}{83.6}        &  57.2      \\ \cline{2-8}

& Codex-10-shot                                                              & \multicolumn{1}{l}{33.3} & \multicolumn{1}{c}{84.6}    &  22.3  & \multicolumn{1}{l}{26.9}     & \multicolumn{1}{c}{76.4}        &  17.3    \\

& Codex-10-shot ($rerank_{token}$)                                            & \multicolumn{1}{l}{32.7} & \multicolumn{1}{c}{\textbf{86.6}}    &  \textbf{27.0}  & \multicolumn{1}{l}{30.9}     & \multicolumn{1}{c}{\textbf{82.4}}        & \textbf{23.2}    
\\

& Codex-10-shot ($rerank_{semantic}$)                                            & \multicolumn{1}{l}{\textbf{35.2}} & \multicolumn{1}{c}{85.6}    & 24.2   & \multicolumn{1}{l}{\textbf{32.8}}     & \multicolumn{1}{c}{81.5}        & 21.6    \\ \cline{2-8} 
& Codex-10-shot ($Selection_{semantic}$)                                                               & \multicolumn{1}{l}{76.3} & \multicolumn{1}{c}{91.2}    & 77.4   & \multicolumn{1}{l}{71.6}     & \multicolumn{1}{c}{85.4}        &  73.6      \\  
& Codex-10-shot ($Selection_{semantic} + Rerank_{token}$)
& \multicolumn{1}{l}{78.8} & \multicolumn{1}{c}{93.5}    & 74.2   & \multicolumn{1}{l}{71.9}     & \multicolumn{1}{c}{85.1}        &   73.9     \\  
& Codex-10-shot ($Selection_{semantic} + Rerank_{semantic}$)
& \multicolumn{1}{l}{\textbf{79.1}} & \multicolumn{1}{c}{\textbf{93.9}}    &  \textbf{75.2}  & \multicolumn{1}{l}{\textbf{72.3}}     & \multicolumn{1}{c}{\textbf{85.7}}        &   \textbf{74.5}      
  
       \\ \hline
\multirow{12}{*}{How-it-is-done}          
& Codex-1-shot                                                               & \multicolumn{1}{l}{21.0} & \multicolumn{1}{c}{39.6}    &  13.5  & \multicolumn{1}{l}{19.1}     & \multicolumn{1}{c}{36.4}        & 12.1       \\  
& Codex-1-shot ($rerank_{token}$)
& \multicolumn{1}{l}{\textbf{29.8}} & \multicolumn{1}{c}{\textbf{79.3}}    & \textbf{22.2}   & \multicolumn{1}{l}{\textbf{27.5}}     & \multicolumn{1}{c}{\textbf{74.8}}        & \textbf{20.9}       \\  
& Codex-1-shot ($rerank_{semantic}$)
& \multicolumn{1}{l}{29.4} & \multicolumn{1}{c}{77.3}    &  21.7  & \multicolumn{1}{l}{26.8}     & \multicolumn{1}{c}{73.1}        &  19.8      \\ \cline{2-8} 

& Codex-1-shot ($Selection_{token}$)                                                               & \multicolumn{1}{l}{31.9} & \multicolumn{1}{c}{72.9}    & 25.8   & \multicolumn{1}{l}{28.6}     & \multicolumn{1}{c}{69.4}        &  24.7      \\  
& Codex-1-shot ($Selection_{token} + Rerank_{token}$)
& \multicolumn{1}{l}{\textbf{33.8}} & \multicolumn{1}{c}{\textbf{79.1}}    &  \textbf{28.9}  & \multicolumn{1}{l}{\textbf{32.2}}     & \multicolumn{1}{c}{\textbf{77.4}}        &  \textbf{26.3}      \\  
& Codex-1-shot ($Selection_{token} + Rerank_{semantic}$)
& \multicolumn{1}{l}{33.0} & \multicolumn{1}{c}{77.6}    &  27.1  & \multicolumn{1}{l}{31.4}     & \multicolumn{1}{c}{75.2}        & 25.8       \\ \cline{2-8}

& Codex-10-shot                                                              & \multicolumn{1}{l}{28.4} & \multicolumn{1}{c}{79.3}    & 19.5   & \multicolumn{1}{l}{21.9}     & \multicolumn{1}{c}{66.7}        &  14.9    \\

& Codex-10-shot ($rerank_{token}$)                                            & \multicolumn{1}{l}{\textbf{30.6}} & \multicolumn{1}{c}{\textbf{95.3}}    &  \textbf{24.7}  & \multicolumn{1}{l}{\textbf{28.1}}     & \multicolumn{1}{c}{\textbf{90.8}}        &  \textbf{23.2}   
\\

& Codex-10-shot ($rerank_{semantic}$)                                            & \multicolumn{1}{l}{30.0} & \multicolumn{1}{c}{95.2}    & 22.3   & \multicolumn{1}{l}{27.6}     & \multicolumn{1}{c}{90.1}        &  20.1   \\ \cline{2-8} 
& Codex-10-shot ($Selection_{semantic}$)                                                               & \multicolumn{1}{l}{51.6} & \multicolumn{1}{c}{86.4}    &  50.8  & \multicolumn{1}{l}{48.9}     & \multicolumn{1}{c}{82.9}        &  47.9      \\  
& Codex-10-shot ($Selection_{semantic} + Rerank_{token}$)
& \multicolumn{1}{l}{\textbf{59.2}} & \multicolumn{1}{c}{\textbf{89.3}}    & \textbf{57.4}   & \multicolumn{1}{l}{\textbf{56.1}}     & \multicolumn{1}{c}{\textbf{85.6}}        & \textbf{53.5}       \\  
& Codex-10-shot ($Selection_{semantic} + Rerank_{semantic}$)
& \multicolumn{1}{l}{57.6} & \multicolumn{1}{c}{88.1}    &  55.2  & \multicolumn{1}{l}{55.3}     & \multicolumn{1}{c}{83.1}        &   53.2      
  
       \\ \hline
\multirow{12}{*}{Property}          
& Codex-1-shot                                                               & \multicolumn{1}{l}{24.7} & \multicolumn{1}{c}{38.4}    &  15.8  & \multicolumn{1}{l}{21.3}     & \multicolumn{1}{c}{33.6}        & 12.4       \\  
& Codex-1-shot ($rerank_{token}$)
& \multicolumn{1}{l}{\textbf{34.8}} & \multicolumn{1}{c}{\textbf{59.8}}    & \textbf{33.2}   & \multicolumn{1}{l}{\textbf{30.6}}     & \multicolumn{1}{c}{\textbf{51.2}}        & \textbf{28.7}       \\  
& Codex-1-shot ($rerank_{semantic}$)
& \multicolumn{1}{l}{34.2} & \multicolumn{1}{c}{59.5}    &  32.1  & \multicolumn{1}{l}{29.5}     & \multicolumn{1}{c}{49.8}        & 27.6       \\ \cline{2-8} 

& Codex-1-shot ($Selection_{token}$)                                                               & \multicolumn{1}{l}{35.7} & \multicolumn{1}{c}{67.7}    & 33.1   & \multicolumn{1}{l}{33.2}     & \multicolumn{1}{c}{64.9}        &  30.8      \\  
& Codex-1-shot ($Selection_{token} + Rerank_{token}$)
& \multicolumn{1}{l}{\textbf{41.6}} & \multicolumn{1}{c}{\textbf{74.2}}    & \textbf{39.2}   & \multicolumn{1}{l}{\textbf{38.4}}     & \multicolumn{1}{c}{\textbf{69.6}}        & \textbf{35.9}       \\  
& Codex-1-shot ($Selection_{token} + Rerank_{semantic}$)
& \multicolumn{1}{l}{40.1} & \multicolumn{1}{c}{72.1}    &  36.0  & \multicolumn{1}{l}{37.7}     & \multicolumn{1}{c}{67.2}        &   34.3     \\ \cline{2-8}

& Codex-10-shot                                                              & \multicolumn{1}{l}{36.2} & \multicolumn{1}{c}{81.9}    & 29.4   & \multicolumn{1}{l}{28.7}     & \multicolumn{1}{c}{80.3}        &  24.7    \\

& Codex-10-shot ($rerank_{token}$)                                            & \multicolumn{1}{l}{\textbf{38.4}} & \multicolumn{1}{c}{\textbf{84.2}}    &  \textbf{31.2}  & \multicolumn{1}{l}{\textbf{35.2}}     & \multicolumn{1}{c}{\textbf{82.7}}        & \textbf{28.6}    
\\

& Codex-10-shot ($rerank_{semantic}$)                                            & \multicolumn{1}{l}{36.2} & \multicolumn{1}{c}{78.4}    & 30.9   & \multicolumn{1}{l}{34.1}     & \multicolumn{1}{c}{81.2}        &  27.4   \\ \cline{2-8} 
& Codex-10-shot ($Selection_{semantic}$)                                                               & \multicolumn{1}{l}{86.3} & \multicolumn{1}{c}{95.8}    & 87.4   & \multicolumn{1}{l}{81.0}     & \multicolumn{1}{c}{86.4}        &  80.8      \\  
& Codex-10-shot ($Selection_{semantic} + Rerank_{token}$)
& \multicolumn{1}{l}{\textbf{87.2}} & \multicolumn{1}{c}{\textbf{96.4}}    &  \textbf{88.7}  & \multicolumn{1}{l}{\textbf{84.9}}     & \multicolumn{1}{c}{\textbf{89.1}}        &  \textbf{83.2}      \\  
& Codex-10-shot ($Selection_{semantic} + Rerank_{semantic}$)
& \multicolumn{1}{l}{86.8} & \multicolumn{1}{c}{96.1}    &  87.9  & \multicolumn{1}{l}{82.5}     & \multicolumn{1}{c}{88.2}        &   82.1      
  
       \\ \hline
\multirow{12}{*}{Average}          
& Codex-1-shot                                                               & \multicolumn{1}{l}{23.1} & \multicolumn{1}{c}{30.7}    &  16.2  & \multicolumn{1}{l}{21.1}     & \multicolumn{1}{c}{26.1}        &  13.6      \\  
& Codex-1-shot ($rerank_{token}$)
& \multicolumn{1}{l}{29.1} & \multicolumn{1}{c}{68.9}    & 25.2   & \multicolumn{1}{l}{26.8}     & \multicolumn{1}{c}{\textbf{64.0}}        &  \textbf{23.2}      \\  
& Codex-1-shot ($rerank_{semantic}$)
& \multicolumn{1}{l}{\textbf{30.2}} & \multicolumn{1}{c}{\textbf{69.0}}    & \textbf{25.8}   & \multicolumn{1}{l}{\textbf{27.2}}     & \multicolumn{1}{c}{63.9}        &  22.2      \\ \cline{2-8} 

& Codex-1-shot ($Selection_{token}$)                                                               & \multicolumn{1}{l}{39.2} & \multicolumn{1}{c}{77.3}    &   35.1 & \multicolumn{1}{l}{36.1}     & \multicolumn{1}{c}{72.8}        &   31.0     \\  
& Codex-1-shot ($Selection_{token} + Rerank_{token}$)
& \multicolumn{1}{l}{\textbf{44.0}} & \multicolumn{1}{c}{\textbf{82.0}}    &  \textbf{41.9}  & \multicolumn{1}{l}{\textbf{41.5}}     & \multicolumn{1}{c}{\textbf{77.4}}        &   \textbf{37.9}     \\  
& Codex-1-shot ($Selection_{token} + Rerank_{semantic}$)
& \multicolumn{1}{l}{43.7} & \multicolumn{1}{c}{80.8}    & 40.9   & \multicolumn{1}{l}{41.4}     & \multicolumn{1}{c}{76.4}        &  37.7      \\ \cline{2-8}

& Codex-10-shot                                                              & \multicolumn{1}{l}{33.4} & \multicolumn{1}{c}{76.1}    & 24.1   & \multicolumn{1}{l}{27.2}     & \multicolumn{1}{c}{66.7}        &  19.2    \\

& Codex-10-shot ($rerank_{token}$)                                            & \multicolumn{1}{l}{35.1} & \multicolumn{1}{c}{\textbf{88.3}}    &  \textbf{27.4}  & \multicolumn{1}{l}{32.0}     & \multicolumn{1}{c}{\textbf{83.8}}        & \textbf{24.5}    
\\

& Codex-10-shot ($rerank_{semantic}$)                                            & \multicolumn{1}{l}{\textbf{35.3}} & \multicolumn{1}{c}{87.1}    & 27.4   & \multicolumn{1}{l}{\textbf{32.4}}     & \multicolumn{1}{c}{83.8}        &   24.2  \\ \cline{2-8} 
& Codex-10-shot ($Selection_{semantic}$)                                                               & \multicolumn{1}{l}{65.9} & \multicolumn{1}{c}{89.4}    & 65.8   & \multicolumn{1}{l}{62.8}     & \multicolumn{1}{c}{83.2}        &   62.7     \\  
& Codex-10-shot ($Selection_{semantic} + Rerank_{token}$)
& \multicolumn{1}{l}{\textbf{72.4}} & \multicolumn{1}{c}{\textbf{91.8}}    &  \textbf{71.6}  & \multicolumn{1}{l}{\textbf{68.8}}     & \multicolumn{1}{c}{\textbf{86.3}}        &  68.6      \\  
& Codex-10-shot ($Selection_{semantic} + Rerank_{semantic}$)
& \multicolumn{1}{l}{72.0} & \multicolumn{1}{c}{91.6}    & 71.5   & \multicolumn{1}{l}{68.4}     & \multicolumn{1}{c}{85.7}        &  \textbf{68.9}       
  
       \\ \hline
\end{tabular}
}
\label{tab:rq3_results}
\end{table*}

%% file: 6.discussion.tex
\section{Discussion}
\label{sec:dis}







\subsection{Human Evaluation}

While metrics such as BLEU, ROUGE-L, and METEOR can evaluate the lexical disparity between the generated comments and the oracle, they are inadequate in reflecting the semantic differences. Thus, to further evaluate the quality of comments generated by various approaches, we conduct a human evaluation.
Specifically, we recruit six participants with at least five years of experience in Java development. The participants include three Ph.D students and three senior researchers who are not co-authors of this paper. 
We randomly select 100 code snippets (20 from each intent category) to perform this user study. For each code snippet, we show the participants the oracle comment and the results from four approaches, namely, DOME, Codex-10-shot, Codex-10-shot with semantic-based selection, and Codex-10-shot with semantic-based selection and token-based reranking, which results in 400 generated comments as our evaluation subjects.
To ensure fairness, the participants are not aware of where the comments are generated from.
Each participant is asked to rate all the 400 comments from three aspects: 
(1) {\bf Naturalness} which reflects the fluency of generated comments from the perspective of grammar;
(2) {\bf Adequacy} which reflects the information richness of generated comments; 
and (3) {\bf Usefulness} which reflects how can generated comments help developers, on a 5-point Likert scale (1 for poor, 2 for marginal, 3 for acceptable, 4 for good, and 5 for excellent).
Such an experiment setting follows existing studies \cite{mu2022automatic,roy2021reassessing}.

Results of our user study are listed in Table~\ref{tab:user_study}.
We observe that higher metric values lead to higher scores rated by the participants.
Specifically, the best-performing model variant in our quantitative evaluation, \ie 10-shot learning with semantic-based demonstration selection and token-based reranking, also achieves the highest scores from participants (\ie 4.3, 4.1, and 3.8 with respect to the three aspects, respectively).
We also note that LLMs are good at generating fluent NL descriptions, since all the model variants achieve scores higher than 4 with respect to the naturalness property.
In contrast, all scores achieved on the usefulness property are lower than 4, which indicates there is still a room for improving the usefulness of the generated comments.

\input{tables/RQ4_human_evaluation}

\subsection{Implications}

{\bf Large language models are few-shot summarizers.}
Our empirical investigation shows that LLMs are capable of generating high-quality code comments with diverse intents.
Results show that the best-performing model variant, \ie Codex-10-shot with semantic-based demonstration selection and token-based reranking, outperforms the state-of-the-art DOME approach to a large extent on the two datasets (\eg outperforms DOME by 128\%/210\% with respect to the BLEU metric on the Funcom/TLC datasets).
This indicates that in practice, developers can refer to the LLMs for helping them automatically generate comments with different intents. LLMs are thus of great potential to facilitate program comprehension activities.
For researchers, this also indicates that the comparison with LLMs is necessary when evaluating a newly-proposed code summarization approach.

{\bf On the importance of prompt quality.}
Our results show that the quality of the prompt provided to LLMs can significantly impact the generated results. 
Specifically, providing LLMs with examples that are similar to the target code may help them generate more qualified results. This calls for more attention to the demonstration selection process.
However, as for the selection strategy, our results also indicate that there is no silver bullet: the token-based similar code selection and the semantic-based one complement each other.
This means that more research efforts could be devoted to devise a better selection strategy.

{\bf More attempts, more gains.}
Due to the sampling process, LLMs can generate multiple results for a specific input. Our results (\eg the case in Figure~\ref{fig:RQ3example}) show that sometimes a comment similar to the oracle one may not be generated at the first place.
Therefore, in practice, developers may query the LLMs for more times if they feel the generated comments are not good enough.
For researchers, how to automatically rerank the results of LLMs also deserves more in-depth explorations and our initial attempt with two simple heuristics achieves promising results.

\subsection{Threats to Validity}
{\bf Internal validity.} 
Codex is trained on open-source projects and thus there may be data leakage, \ie Codex may have seen the comments for the test cases during its pre-training.
However, we observe that Codex does not perform effectively under the zero-shot setting, which indicates that the model's output is not generated due to memorization.
Such a threat is also faced by other studies on large language models \cite{nashid2023retrieval}, and to fully address this threat requires to re-train the model from scratch, which would be currently infeasible considering the limitation of the computation resource.

As introduced, our results are affected by the randomness incurred by the model sampling process or the demonstration selection. To mitigate this threat as well as keep the time cost of the experiments in a reasonable scale, we repeat each experiment for one hundred times.
However, one hundred may not fully eliminate the randomness and we leave more experiments as future work.

{\bf External validity.}
The first threat to applying our observation in practice is that it is unclear whether developers can find code snippets similar to the target code for constructing better prompts to the LLMs. 
However, our results also show that under the 10-shot setting, the performance of Codex exceeds that of the state-of-the-art DOME even if the demonstrations are randomly selected.

Another threat is that we only focus on Java programming language. This setting is restricted by the availability of multi-intent comment dataset in the literature.
This threat is alleviated considering that the two datasets are large-scale and Java is the most widely-studied language in the comment generation domain \cite{hu2018deep,mu2023developer,leclair2019neural}.

%% file: tables/RQ4_human_evaluation.tex
\begin{table}[!t]
\centering
\caption{The statistic results of our user study.}
\label{tab:user_study}
\resizebox{0.9\linewidth}{!}{
\begin{tabular}{c|l|c|c}
\hline & \multicolumn{1}{c|}{Approach} & Avg.  & Std. \\
\hline \multirow{4}{*}{ Naturalness } & DOME & 3.9  & 0.8   \\
& Codex-10-shot & 4.2 & 0.7   \\
& Codex-10-shot ($Selection_{semantic}$) & 4.3  & 0.8 \\
& Codex-10-shot ($Selection_{semantic}$+$Rerank_{token}$) & \textbf{4.3} & 0.7   \\

\hline \multirow{4}{*}{ Adequacy } & DOME & 3.3  & 1.3  \\
& Codex-10-shot & 3.5  & 1.1  \\
& Codex-10-shot ($Selection_{semantic}$)   & 3.8 & 1.2 \\
& Codex-10-shot ($Selection_{semantic}$+$Rerank_{token}$) & \textbf{4.1} & 0.9   \\
\hline \multirow{4}{*}{ Usefulness } & DOME & 3.0  &  1.4  \\
& Codex-10-shot & 3.1 & 1.3   \\
& Codex-10-shot ($Selection_{semantic}$)   & 3.7 & 1.1 \\
& Codex-10-shot ($Selection_{semantic}$+$Rerank_{token}$) & \textbf{3.8} & 1.3   \\
\hline
\end{tabular}
}
\end{table}

%% file: 7.relatedwork.tex


%% file: 8.conclusion.tex
\section{Conclusion}
\label{sec:conc}

Our empirical study mainly investigates whether it is feasible to utilize the LLMs for addressing multi-intent comment generation and further how to improve the effectiveness of LLMs on this task.
Our results gives positive answer to the first point: by utilizing few-shot in-context learning, the performance of Codex exceeds that of the state-of-the-art supervised learning approach.
We also demonstrate that both demonstration selection and result reranking can help boost the performance of Codex.
Our study establishes new baselines for the multi-intent comment generation task as well as pointing research directions that deserve more in-depth investigations.

\section{Data Availability}
All code and data in this study are publicly available at:
\begin{center}
   {\bf \url{https://github.com/gmy2013/LLM_Comment_Generation}}.
\end{center}

%% file: main.bbl
\begin{thebibliography}{78}


\ifx \showCODEN    \undefined \def \showCODEN     #1{\unskip}     \fi
\ifx \showDOI      \undefined \def \showDOI       #1{#1}\fi
\ifx \showISBNx    \undefined \def \showISBNx     #1{\unskip}     \fi
\ifx \showISBNxiii \undefined \def \showISBNxiii  #1{\unskip}     \fi
\ifx \showISSN     \undefined \def \showISSN      #1{\unskip}     \fi
\ifx \showLCCN     \undefined \def \showLCCN      #1{\unskip}     \fi
\ifx \shownote     \undefined \def \shownote      #1{#1}          \fi
\ifx \showarticletitle \undefined \def \showarticletitle #1{#1}   \fi
\ifx \showURL      \undefined \def \showURL       {\relax}        \fi
\providecommand\bibfield[2]{#2}
\providecommand\bibinfo[2]{#2}
\providecommand\natexlab[1]{#1}
\providecommand\showeprint[2][]{arXiv:#2}

\bibitem[\protect\citeauthoryear{??}{ref}{2010}]%
        {ref:funcom}
 \bibinfo{year}{2010}\natexlab{}.
\newblock \bibinfo{title}{Original Funcom Dataset}.  (\bibinfo{year}{2010}).
\newblock
\newblock
\shownote{http://www.ics.uci.edu/lopes/datasets/.}


\bibitem[\protect\citeauthoryear{??}{ref}{2022}]%
        {ref:codex}
 \bibinfo{year}{2022}\natexlab{}.
\newblock \bibinfo{title}{Codex model}.  (\bibinfo{year}{2022}).
\newblock
\newblock
\shownote{https://beta.openai.com/docs/models/codex-seriesprivate-beta.}


\bibitem[\protect\citeauthoryear{Ahmad, Chakraborty, Ray, and Chang}{Ahmad
  et~al\mbox{.}}{2020}]%
        {ahmad2020transformer}
\bibfield{author}{\bibinfo{person}{Wasi~Uddin Ahmad}, \bibinfo{person}{Saikat
  Chakraborty}, \bibinfo{person}{Baishakhi Ray}, {and} \bibinfo{person}{Kai-Wei
  Chang}.} \bibinfo{year}{2020}\natexlab{}.
\newblock \showarticletitle{A transformer-based approach for source code
  summarization}.
\newblock \bibinfo{journal}{\emph{arXiv preprint arXiv:2005.00653}}
  (\bibinfo{year}{2020}).
\newblock


\bibitem[\protect\citeauthoryear{Alon, Brody, Levy, and Yahav}{Alon
  et~al\mbox{.}}{2018}]%
        {alon2018code2seq}
\bibfield{author}{\bibinfo{person}{Uri Alon}, \bibinfo{person}{Shaked Brody},
  \bibinfo{person}{Omer Levy}, {and} \bibinfo{person}{Eran Yahav}.}
  \bibinfo{year}{2018}\natexlab{}.
\newblock \showarticletitle{code2seq: Generating sequences from structured
  representations of code}.
\newblock \bibinfo{journal}{\emph{arXiv preprint arXiv:1808.01400}}
  (\bibinfo{year}{2018}).
\newblock


\bibitem[\protect\citeauthoryear{Alon, Brody, Levy, and Yahav}{Alon
  et~al\mbox{.}}{2019}]%
        {alon2019code2seq}
\bibfield{author}{\bibinfo{person}{Uri Alon}, \bibinfo{person}{Shaked Brody},
  \bibinfo{person}{Omer Levy}, {and} \bibinfo{person}{Eran Yahav}.}
  \bibinfo{year}{2019}\natexlab{}.
\newblock \showarticletitle{code2seq: Generating Sequences from Structured
  Representations of Code}. In \bibinfo{booktitle}{\emph{Proceedings of the 7th
  International Conference on Learning Representations}}.
  \bibinfo{publisher}{OpenReview.net}.
\newblock


\bibitem[\protect\citeauthoryear{Banerjee and Lavie}{Banerjee and
  Lavie}{2005}]%
        {banerjee2005meteor}
\bibfield{author}{\bibinfo{person}{Satanjeev Banerjee} {and}
  \bibinfo{person}{Alon Lavie}.} \bibinfo{year}{2005}\natexlab{}.
\newblock \showarticletitle{METEOR: An automatic metric for MT evaluation with
  improved correlation with human judgments}. In
  \bibinfo{booktitle}{\emph{Proceedings of the acl workshop on intrinsic and
  extrinsic evaluation measures for machine translation and/or summarization}}.
  \bibinfo{pages}{65--72}.
\newblock


\bibitem[\protect\citeauthoryear{Bansal, Haque, and McMillan}{Bansal
  et~al\mbox{.}}{2021}]%
        {bansal2021project}
\bibfield{author}{\bibinfo{person}{Aakash Bansal}, \bibinfo{person}{Sakib
  Haque}, {and} \bibinfo{person}{Collin McMillan}.}
  \bibinfo{year}{2021}\natexlab{}.
\newblock \showarticletitle{Project-level encoding for neural source code
  summarization of subroutines}. In \bibinfo{booktitle}{\emph{2021 IEEE/ACM
  29th International Conference on Program Comprehension (ICPC)}}. IEEE,
  \bibinfo{pages}{253--264}.
\newblock


\bibitem[\protect\citeauthoryear{Brown, Mann, Ryder, Subbiah, Kaplan, Dhariwal,
  Neelakantan, Shyam, Sastry, Askell, et~al\mbox{.}}{Brown
  et~al\mbox{.}}{2020}]%
        {brown2020language}
\bibfield{author}{\bibinfo{person}{Tom Brown}, \bibinfo{person}{Benjamin Mann},
  \bibinfo{person}{Nick Ryder}, \bibinfo{person}{Melanie Subbiah},
  \bibinfo{person}{Jared~D Kaplan}, \bibinfo{person}{Prafulla Dhariwal},
  \bibinfo{person}{Arvind Neelakantan}, \bibinfo{person}{Pranav Shyam},
  \bibinfo{person}{Girish Sastry}, \bibinfo{person}{Amanda Askell},
  {et~al\mbox{.}}} \bibinfo{year}{2020}\natexlab{}.
\newblock \showarticletitle{Language models are few-shot learners}.
\newblock \bibinfo{journal}{\emph{Advances in neural information processing
  systems}}  \bibinfo{volume}{33} (\bibinfo{year}{2020}),
  \bibinfo{pages}{1877--1901}.
\newblock


\bibitem[\protect\citeauthoryear{Cai, Liang, Xu, Li, Hao, and Chen}{Cai
  et~al\mbox{.}}{2020}]%
        {cai2020tag}
\bibfield{author}{\bibinfo{person}{Ruichu Cai}, \bibinfo{person}{Zhihao Liang},
  \bibinfo{person}{Boyan Xu}, \bibinfo{person}{Zijian Li},
  \bibinfo{person}{Yuexing Hao}, {and} \bibinfo{person}{Yao Chen}.}
  \bibinfo{year}{2020}\natexlab{}.
\newblock \showarticletitle{TAG: Type auxiliary guiding for code comment
  generation}.
\newblock \bibinfo{journal}{\emph{arXiv preprint arXiv:2005.02835}}
  (\bibinfo{year}{2020}).
\newblock


\bibitem[\protect\citeauthoryear{Chen, Zhang, Nguyen, Zan, Lin, Lou, and
  Chen}{Chen et~al\mbox{.}}{2022}]%
        {chen2022codet}
\bibfield{author}{\bibinfo{person}{Bei Chen}, \bibinfo{person}{Fengji Zhang},
  \bibinfo{person}{Anh Nguyen}, \bibinfo{person}{Daoguang Zan},
  \bibinfo{person}{Zeqi Lin}, \bibinfo{person}{Jian-Guang Lou}, {and}
  \bibinfo{person}{Weizhu Chen}.} \bibinfo{year}{2022}\natexlab{}.
\newblock \showarticletitle{Codet: Code generation with generated tests}.
\newblock \bibinfo{journal}{\emph{arXiv preprint arXiv:2207.10397}}
  (\bibinfo{year}{2022}).
\newblock


\bibitem[\protect\citeauthoryear{Chen, Tworek, Jun, Yuan, Pinto, Kaplan,
  Edwards, Burda, Joseph, Brockman, et~al\mbox{.}}{Chen et~al\mbox{.}}{2021a}]%
        {chen2021evaluating}
\bibfield{author}{\bibinfo{person}{Mark Chen}, \bibinfo{person}{Jerry Tworek},
  \bibinfo{person}{Heewoo Jun}, \bibinfo{person}{Qiming Yuan},
  \bibinfo{person}{Henrique Ponde de~Oliveira Pinto}, \bibinfo{person}{Jared
  Kaplan}, \bibinfo{person}{Harri Edwards}, \bibinfo{person}{Yuri Burda},
  \bibinfo{person}{Nicholas Joseph}, \bibinfo{person}{Greg Brockman},
  {et~al\mbox{.}}} \bibinfo{year}{2021}\natexlab{a}.
\newblock \showarticletitle{Evaluating large language models trained on code}.
\newblock \bibinfo{journal}{\emph{arXiv preprint arXiv:2107.03374}}
  (\bibinfo{year}{2021}).
\newblock


\bibitem[\protect\citeauthoryear{Chen, Xia, Hu, Lo, and Li}{Chen
  et~al\mbox{.}}{2021b}]%
        {chen2021my}
\bibfield{author}{\bibinfo{person}{Qiuyuan Chen}, \bibinfo{person}{Xin Xia},
  \bibinfo{person}{Han Hu}, \bibinfo{person}{David Lo}, {and}
  \bibinfo{person}{Shanping Li}.} \bibinfo{year}{2021}\natexlab{b}.
\newblock \showarticletitle{Why my code summarization model does not work: Code
  comment improvement with category prediction}.
\newblock \bibinfo{journal}{\emph{ACM Transactions on Software Engineering and
  Methodology (TOSEM)}} \bibinfo{volume}{30}, \bibinfo{number}{2}
  (\bibinfo{year}{2021}), \bibinfo{pages}{1--29}.
\newblock


\bibitem[\protect\citeauthoryear{Cheng, Fostiropoulos, and Boehm}{Cheng
  et~al\mbox{.}}{2021}]%
        {cheng2021gn}
\bibfield{author}{\bibinfo{person}{Junyan Cheng}, \bibinfo{person}{Iordanis
  Fostiropoulos}, {and} \bibinfo{person}{Barry Boehm}.}
  \bibinfo{year}{2021}\natexlab{}.
\newblock \showarticletitle{GN-Transformer: Fusing Sequence and Graph
  Representation for Improved Code Summarization}.
\newblock \bibinfo{journal}{\emph{arXiv preprint arXiv:2111.08874}}
  (\bibinfo{year}{2021}).
\newblock


\bibitem[\protect\citeauthoryear{Dakhel, Majdinasab, Nikanjam, Khomh,
  Desmarais, Ming, et~al\mbox{.}}{Dakhel et~al\mbox{.}}{2022}]%
        {dakhel2022github}
\bibfield{author}{\bibinfo{person}{Arghavan~Moradi Dakhel},
  \bibinfo{person}{Vahid Majdinasab}, \bibinfo{person}{Amin Nikanjam},
  \bibinfo{person}{Foutse Khomh}, \bibinfo{person}{Michel~C Desmarais},
  \bibinfo{person}{Zhen Ming}, {et~al\mbox{.}}}
  \bibinfo{year}{2022}\natexlab{}.
\newblock \showarticletitle{GitHub Copilot AI pair programmer: Asset or
  Liability?}
\newblock \bibinfo{journal}{\emph{arXiv preprint arXiv:2206.15331}}
  (\bibinfo{year}{2022}).
\newblock


\bibitem[\protect\citeauthoryear{Devlin, Chang, Lee, and Toutanova}{Devlin
  et~al\mbox{.}}{2019}]%
        {devlin2019bert}
\bibfield{author}{\bibinfo{person}{Jacob Devlin}, \bibinfo{person}{Ming{-}Wei
  Chang}, \bibinfo{person}{Kenton Lee}, {and} \bibinfo{person}{Kristina
  Toutanova}.} \bibinfo{year}{2019}\natexlab{}.
\newblock \showarticletitle{{BERT:} Pre-training of Deep Bidirectional
  Transformers for Language Understanding}. In
  \bibinfo{booktitle}{\emph{Proceedings of the 2019 Conference of the North
  American Chapter of the Association for Computational Linguistics: Human
  Language Technologies}}. \bibinfo{pages}{4171--4186}.
\newblock
\urldef\tempurl%
\url{https://doi.org/10.18653/v1/n19-1423}
\showDOI{\tempurl}


\bibitem[\protect\citeauthoryear{Devlin, Chang, Lee, and Toutanova}{Devlin
  et~al\mbox{.}}{2018}]%
        {devlin2018bert}
\bibfield{author}{\bibinfo{person}{Jacob Devlin}, \bibinfo{person}{Ming-Wei
  Chang}, \bibinfo{person}{Kenton Lee}, {and} \bibinfo{person}{Kristina
  Toutanova}.} \bibinfo{year}{2018}\natexlab{}.
\newblock \showarticletitle{Bert: Pre-training of deep bidirectional
  transformers for language understanding}.
\newblock \bibinfo{journal}{\emph{arXiv preprint arXiv:1810.04805}}
  (\bibinfo{year}{2018}).
\newblock


\bibitem[\protect\citeauthoryear{Eghbali and Pradel}{Eghbali and
  Pradel}{2022}]%
        {eghbali2022crystalbleu}
\bibfield{author}{\bibinfo{person}{Aryaz Eghbali} {and}
  \bibinfo{person}{Michael Pradel}.} \bibinfo{year}{2022}\natexlab{}.
\newblock \showarticletitle{CrystalBLEU: precisely and efficiently measuring
  the similarity of code}. In \bibinfo{booktitle}{\emph{37th IEEE/ACM
  International Conference on Automated Software Engineering}}.
  \bibinfo{pages}{1--12}.
\newblock


\bibitem[\protect\citeauthoryear{Fan, Gao, Roychoudhury, and Tan}{Fan
  et~al\mbox{.}}{2022}]%
        {fan2022improving}
\bibfield{author}{\bibinfo{person}{Zhiyu Fan}, \bibinfo{person}{Xiang Gao},
  \bibinfo{person}{Abhik Roychoudhury}, {and} \bibinfo{person}{Shin~Hwei Tan}.}
  \bibinfo{year}{2022}\natexlab{}.
\newblock \showarticletitle{Improving automatically generated code from Codex
  via Automated Program Repair}.
\newblock \bibinfo{journal}{\emph{arXiv preprint arXiv:2205.10583}}
  (\bibinfo{year}{2022}).
\newblock


\bibitem[\protect\citeauthoryear{Feng, Guo, Tang, Duan, Feng, Gong, Shou, Qin,
  Liu, Jiang, et~al\mbox{.}}{Feng et~al\mbox{.}}{2020}]%
        {feng2020codebert}
\bibfield{author}{\bibinfo{person}{Zhangyin Feng}, \bibinfo{person}{Daya Guo},
  \bibinfo{person}{Duyu Tang}, \bibinfo{person}{Nan Duan},
  \bibinfo{person}{Xiaocheng Feng}, \bibinfo{person}{Ming Gong},
  \bibinfo{person}{Linjun Shou}, \bibinfo{person}{Bing Qin},
  \bibinfo{person}{Ting Liu}, \bibinfo{person}{Daxin Jiang}, {et~al\mbox{.}}}
  \bibinfo{year}{2020}\natexlab{}.
\newblock \showarticletitle{Codebert: A pre-trained model for programming and
  natural languages}.
\newblock \bibinfo{journal}{\emph{arXiv preprint arXiv:2002.08155}}
  (\bibinfo{year}{2020}).
\newblock


\bibitem[\protect\citeauthoryear{Gao, Gao, He, Zeng, Nie, Xia, and Lyu}{Gao
  et~al\mbox{.}}{2023}]%
        {gao2023code}
\bibfield{author}{\bibinfo{person}{Shuzheng Gao}, \bibinfo{person}{Cuiyun Gao},
  \bibinfo{person}{Yulan He}, \bibinfo{person}{Jichuan Zeng},
  \bibinfo{person}{Lunyiu Nie}, \bibinfo{person}{Xin Xia}, {and}
  \bibinfo{person}{Michael Lyu}.} \bibinfo{year}{2023}\natexlab{}.
\newblock \showarticletitle{Code Structure--Guided Transformer for Source Code
  Summarization}.
\newblock \bibinfo{journal}{\emph{ACM Transactions on Software Engineering and
  Methodology}} \bibinfo{volume}{32}, \bibinfo{number}{1}
  (\bibinfo{year}{2023}), \bibinfo{pages}{1--32}.
\newblock


\bibitem[\protect\citeauthoryear{Geng, Wang, Dong, Gu, Peng, Ruan, and
  Liao}{Geng et~al\mbox{.}}{2022}]%
        {geng2022fine}
\bibfield{author}{\bibinfo{person}{Mingyang Geng}, \bibinfo{person}{Shangwen
  Wang}, \bibinfo{person}{Dezun Dong}, \bibinfo{person}{Shanzhi Gu},
  \bibinfo{person}{Fang Peng}, \bibinfo{person}{Weijian Ruan}, {and}
  \bibinfo{person}{Xiangke Liao}.} \bibinfo{year}{2022}\natexlab{}.
\newblock \showarticletitle{Fine-grained code-comment semantic interaction
  analysis}. In \bibinfo{booktitle}{\emph{Proceedings of the 30th IEEE/ACM
  International Conference on Program Comprehension}}.
  \bibinfo{pages}{585--596}.
\newblock


\bibitem[\protect\citeauthoryear{Geng, Wang, Dong, Wang, Cao, Zhang, and
  Jin}{Geng et~al\mbox{.}}{2023}]%
        {geng2023interpretation}
\bibfield{author}{\bibinfo{person}{Mingyang Geng}, \bibinfo{person}{Shangwen
  Wang}, \bibinfo{person}{Dezun Dong}, \bibinfo{person}{Haotian Wang},
  \bibinfo{person}{Shaomeng Cao}, \bibinfo{person}{Kechi Zhang}, {and}
  \bibinfo{person}{Zhi Jin}.} \bibinfo{year}{2023}\natexlab{}.
\newblock \showarticletitle{Interpretation-based Code Summarization}. In
  \bibinfo{booktitle}{\emph{Proceedings of the 31st IEEE/ACM International
  Conference on Program Comprehension}}.
\newblock


\bibitem[\protect\citeauthoryear{Golubev, Poletansky, Povarov, and
  Bryksin}{Golubev et~al\mbox{.}}{2021}]%
        {golubev2021multi}
\bibfield{author}{\bibinfo{person}{Yaroslav Golubev}, \bibinfo{person}{Viktor
  Poletansky}, \bibinfo{person}{Nikita Povarov}, {and} \bibinfo{person}{Timofey
  Bryksin}.} \bibinfo{year}{2021}\natexlab{}.
\newblock \showarticletitle{Multi-threshold token-based code clone detection}.
  In \bibinfo{booktitle}{\emph{2021 IEEE International Conference on Software
  Analysis, Evolution and Reengineering (SANER)}}. IEEE,
  \bibinfo{pages}{496--500}.
\newblock


\bibitem[\protect\citeauthoryear{Guo, Ren, Lu, Feng, Tang, Liu, Zhou, Duan,
  Svyatkovskiy, Fu, et~al\mbox{.}}{Guo et~al\mbox{.}}{2021}]%
        {guo2021graphcodebert}
\bibfield{author}{\bibinfo{person}{Daya Guo}, \bibinfo{person}{Shuo Ren},
  \bibinfo{person}{Shuai Lu}, \bibinfo{person}{Zhangyin Feng},
  \bibinfo{person}{Duyu Tang}, \bibinfo{person}{Shujie Liu},
  \bibinfo{person}{Long Zhou}, \bibinfo{person}{Nan Duan},
  \bibinfo{person}{Alexey Svyatkovskiy}, \bibinfo{person}{Shengyu Fu},
  {et~al\mbox{.}}} \bibinfo{year}{2021}\natexlab{}.
\newblock \showarticletitle{GraphCodeBERT: Pre-training Code Representations
  with Data Flow}. In \bibinfo{booktitle}{\emph{ICLR}}.
\newblock


\bibitem[\protect\citeauthoryear{Haiduc, Aponte, Moreno, and Marcus}{Haiduc
  et~al\mbox{.}}{2010}]%
        {haiduc2010use}
\bibfield{author}{\bibinfo{person}{Sonia Haiduc}, \bibinfo{person}{Jairo
  Aponte}, \bibinfo{person}{Laura Moreno}, {and} \bibinfo{person}{Andrian
  Marcus}.} \bibinfo{year}{2010}\natexlab{}.
\newblock \showarticletitle{On the use of automated text summarization
  techniques for summarizing source code}. In \bibinfo{booktitle}{\emph{2010
  17th Working Conference on Reverse Engineering}}. IEEE,
  \bibinfo{pages}{35--44}.
\newblock


\bibitem[\protect\citeauthoryear{Haque, LeClair, Wu, and McMillan}{Haque
  et~al\mbox{.}}{2020}]%
        {haque2020improved}
\bibfield{author}{\bibinfo{person}{Sakib Haque}, \bibinfo{person}{Alexander
  LeClair}, \bibinfo{person}{Lingfei Wu}, {and} \bibinfo{person}{Collin
  McMillan}.} \bibinfo{year}{2020}\natexlab{}.
\newblock \showarticletitle{Improved automatic summarization of subroutines via
  attention to file context}. In \bibinfo{booktitle}{\emph{Proceedings of the
  17th International Conference on Mining Software Repositories}}.
  \bibinfo{pages}{300--310}.
\newblock


\bibitem[\protect\citeauthoryear{Hill, Pollock, and Vijay-Shanker}{Hill
  et~al\mbox{.}}{2009}]%
        {hill2009automatically}
\bibfield{author}{\bibinfo{person}{Emily Hill}, \bibinfo{person}{Lori Pollock},
  {and} \bibinfo{person}{K Vijay-Shanker}.} \bibinfo{year}{2009}\natexlab{}.
\newblock \showarticletitle{Automatically capturing source code context of
  nl-queries for software maintenance and reuse}. In
  \bibinfo{booktitle}{\emph{2009 IEEE 31st International Conference on Software
  Engineering}}. IEEE, \bibinfo{pages}{232--242}.
\newblock


\bibitem[\protect\citeauthoryear{Hu, Li, Xia, Lo, and Jin}{Hu
  et~al\mbox{.}}{2018a}]%
        {hu2018deep}
\bibfield{author}{\bibinfo{person}{Xing Hu}, \bibinfo{person}{Ge Li},
  \bibinfo{person}{Xin Xia}, \bibinfo{person}{David Lo}, {and}
  \bibinfo{person}{Zhi Jin}.} \bibinfo{year}{2018}\natexlab{a}.
\newblock \showarticletitle{Deep code comment generation}. In
  \bibinfo{booktitle}{\emph{Proceedings of the 26th conference on program
  comprehension}}. \bibinfo{pages}{200--210}.
\newblock


\bibitem[\protect\citeauthoryear{Hu, Li, Xia, Lo, and Jin}{Hu
  et~al\mbox{.}}{2020}]%
        {hu2020deep}
\bibfield{author}{\bibinfo{person}{Xing Hu}, \bibinfo{person}{Ge Li},
  \bibinfo{person}{Xin Xia}, \bibinfo{person}{David Lo}, {and}
  \bibinfo{person}{Zhi Jin}.} \bibinfo{year}{2020}\natexlab{}.
\newblock \showarticletitle{Deep code comment generation with hybrid lexical
  and syntactical information}.
\newblock \bibinfo{journal}{\emph{Empirical Software Engineering}}
  \bibinfo{volume}{25} (\bibinfo{year}{2020}), \bibinfo{pages}{2179--2217}.
\newblock


\bibitem[\protect\citeauthoryear{Hu, Li, Xia, Lo, Lu, and Jin}{Hu
  et~al\mbox{.}}{2018b}]%
        {hu2018summarizing}
\bibfield{author}{\bibinfo{person}{Xing Hu}, \bibinfo{person}{Ge Li},
  \bibinfo{person}{Xin Xia}, \bibinfo{person}{David Lo}, \bibinfo{person}{Shuai
  Lu}, {and} \bibinfo{person}{Zhi Jin}.} \bibinfo{year}{2018}\natexlab{b}.
\newblock \showarticletitle{Summarizing source code with transferred api
  knowledge}.
\newblock  (\bibinfo{year}{2018}).
\newblock


\bibitem[\protect\citeauthoryear{Husain, Wu, Gazit, Allamanis, and
  Brockschmidt}{Husain et~al\mbox{.}}{2019}]%
        {husain2019codesearchnet}
\bibfield{author}{\bibinfo{person}{Hamel Husain}, \bibinfo{person}{Ho-Hsiang
  Wu}, \bibinfo{person}{Tiferet Gazit}, \bibinfo{person}{Miltiadis Allamanis},
  {and} \bibinfo{person}{Marc Brockschmidt}.} \bibinfo{year}{2019}\natexlab{}.
\newblock \showarticletitle{Codesearchnet challenge: Evaluating the state of
  semantic code search}.
\newblock \bibinfo{journal}{\emph{arXiv preprint arXiv:1909.09436}}
  (\bibinfo{year}{2019}).
\newblock


\bibitem[\protect\citeauthoryear{Iyer, Konstas, Cheung, and Zettlemoyer}{Iyer
  et~al\mbox{.}}{2016}]%
        {iyer2016summarizing}
\bibfield{author}{\bibinfo{person}{Srinivasan Iyer}, \bibinfo{person}{Ioannis
  Konstas}, \bibinfo{person}{Alvin Cheung}, {and} \bibinfo{person}{Luke
  Zettlemoyer}.} \bibinfo{year}{2016}\natexlab{}.
\newblock \showarticletitle{Summarizing source code using a neural attention
  model}. In \bibinfo{booktitle}{\emph{Proceedings of the 54th Annual Meeting
  of the Association for Computational Linguistics (Volume 1: Long Papers)}}.
  \bibinfo{pages}{2073--2083}.
\newblock


\bibitem[\protect\citeauthoryear{Kamiya, Kusumoto, and Inoue}{Kamiya
  et~al\mbox{.}}{2002}]%
        {kamiya2002ccfinder}
\bibfield{author}{\bibinfo{person}{Toshihiro Kamiya}, \bibinfo{person}{Shinji
  Kusumoto}, {and} \bibinfo{person}{Katsuro Inoue}.}
  \bibinfo{year}{2002}\natexlab{}.
\newblock \showarticletitle{CCFinder: A multilinguistic token-based code clone
  detection system for large scale source code}.
\newblock \bibinfo{journal}{\emph{IEEE transactions on software engineering}}
  \bibinfo{volume}{28}, \bibinfo{number}{7} (\bibinfo{year}{2002}),
  \bibinfo{pages}{654--670}.
\newblock


\bibitem[\protect\citeauthoryear{Kolak, Martins, Le~Goues, and
  Hellendoorn}{Kolak et~al\mbox{.}}{2022}]%
        {kolak2022patch}
\bibfield{author}{\bibinfo{person}{Sophia~D Kolak}, \bibinfo{person}{Ruben
  Martins}, \bibinfo{person}{Claire Le~Goues}, {and}
  \bibinfo{person}{Vincent~Josua Hellendoorn}.}
  \bibinfo{year}{2022}\natexlab{}.
\newblock \showarticletitle{Patch Generation with Language Models: Feasibility
  and Scaling Behavior}. In \bibinfo{booktitle}{\emph{Deep Learning for Code
  Workshop}}.
\newblock


\bibitem[\protect\citeauthoryear{LeClair, Bansal, and McMillan}{LeClair
  et~al\mbox{.}}{2021}]%
        {leclair2021ensemble}
\bibfield{author}{\bibinfo{person}{Alexander LeClair}, \bibinfo{person}{Aakash
  Bansal}, {and} \bibinfo{person}{Collin McMillan}.}
  \bibinfo{year}{2021}\natexlab{}.
\newblock \showarticletitle{Ensemble models for neural source code
  summarization of subroutines}. In \bibinfo{booktitle}{\emph{2021 IEEE
  International Conference on Software Maintenance and Evolution (ICSME)}}.
  IEEE, \bibinfo{pages}{286--297}.
\newblock


\bibitem[\protect\citeauthoryear{LeClair, Jiang, and McMillan}{LeClair
  et~al\mbox{.}}{2019}]%
        {leclair2019neural}
\bibfield{author}{\bibinfo{person}{Alexander LeClair}, \bibinfo{person}{Siyuan
  Jiang}, {and} \bibinfo{person}{Collin McMillan}.}
  \bibinfo{year}{2019}\natexlab{}.
\newblock \showarticletitle{A neural model for generating natural language
  summaries of program subroutines}. In \bibinfo{booktitle}{\emph{2019 IEEE/ACM
  41st International Conference on Software Engineering (ICSE)}}. IEEE,
  \bibinfo{pages}{795--806}.
\newblock


\bibitem[\protect\citeauthoryear{Li, Li, Li, Hu, Xia, and Jin}{Li
  et~al\mbox{.}}{2021}]%
        {li2021editsum}
\bibfield{author}{\bibinfo{person}{Jia Li}, \bibinfo{person}{Yongmin Li},
  \bibinfo{person}{Ge Li}, \bibinfo{person}{Xing Hu}, \bibinfo{person}{Xin
  Xia}, {and} \bibinfo{person}{Zhi Jin}.} \bibinfo{year}{2021}\natexlab{}.
\newblock \showarticletitle{Editsum: A retrieve-and-edit framework for source
  code summarization}. In \bibinfo{booktitle}{\emph{2021 36th IEEE/ACM
  International Conference on Automated Software Engineering (ASE)}}. IEEE,
  \bibinfo{pages}{155--166}.
\newblock


\bibitem[\protect\citeauthoryear{Li, Yang, Jiang, Yan, Luo, Hua, Liang, and
  Zuo}{Li et~al\mbox{.}}{2022}]%
        {li2022auger}
\bibfield{author}{\bibinfo{person}{Lingwei Li}, \bibinfo{person}{Li Yang},
  \bibinfo{person}{Huaxi Jiang}, \bibinfo{person}{Jun Yan},
  \bibinfo{person}{Tiejian Luo}, \bibinfo{person}{Zihan Hua},
  \bibinfo{person}{Geng Liang}, {and} \bibinfo{person}{Chun Zuo}.}
  \bibinfo{year}{2022}\natexlab{}.
\newblock \showarticletitle{AUGER: automatically generating review comments
  with pre-training models}. In \bibinfo{booktitle}{\emph{Proceedings of the
  30th ACM Joint European Software Engineering Conference and Symposium on the
  Foundations of Software Engineering}}. \bibinfo{pages}{1009--1021}.
\newblock


\bibitem[\protect\citeauthoryear{Lin, Wang, Liu, Mao, and Bissyand{\'e}}{Lin
  et~al\mbox{.}}{2021b}]%
        {lin2021automated}
\bibfield{author}{\bibinfo{person}{Bo Lin}, \bibinfo{person}{Shangwen Wang},
  \bibinfo{person}{Kui Liu}, \bibinfo{person}{Xiaoguang Mao}, {and}
  \bibinfo{person}{Tegawend{\'e}~F. Bissyand{\'e}}.}
  \bibinfo{year}{2021}\natexlab{b}.
\newblock \showarticletitle{Automated Comment Update: How Far are We?}. In
  \bibinfo{booktitle}{\emph{Proceedings of the 29th IEEE/ACM International
  Conference on Program Comprehension (ICPC)}}. \bibinfo{pages}{36--46}.
\newblock
\urldef\tempurl%
\url{https://doi.org/10.1109/ICPC52881.2021.00013}
\showDOI{\tempurl}


\bibitem[\protect\citeauthoryear{Lin, Wang, Liu, Xia, and Mao}{Lin
  et~al\mbox{.}}{2023}]%
        {lin2022predictive}
\bibfield{author}{\bibinfo{person}{Bo Lin}, \bibinfo{person}{Shangwen Wang},
  \bibinfo{person}{Zhongxin Liu}, \bibinfo{person}{Xin Xia}, {and}
  \bibinfo{person}{Xiaoguang Mao}.} \bibinfo{year}{2023}\natexlab{}.
\newblock \showarticletitle{Predictive Comment Updating With Heuristics and
  AST-Path-Based Neural Learning: A Two-Phase Approach}.
\newblock \bibinfo{journal}{\emph{IEEE Transactions on Software Engineering}}
  \bibinfo{volume}{49}, \bibinfo{number}{4} (\bibinfo{year}{2023}),
  \bibinfo{pages}{1640--1660}.
\newblock
\urldef\tempurl%
\url{https://doi.org/10.1109/TSE.2022.3185458}
\showDOI{\tempurl}


\bibitem[\protect\citeauthoryear{Lin, Ouyang, Zhuang, Chen, Li, and Wu}{Lin
  et~al\mbox{.}}{2021a}]%
        {lin2021improving}
\bibfield{author}{\bibinfo{person}{Chen Lin}, \bibinfo{person}{Zhichao Ouyang},
  \bibinfo{person}{Junqing Zhuang}, \bibinfo{person}{Jianqiang Chen},
  \bibinfo{person}{Hui Li}, {and} \bibinfo{person}{Rongxin Wu}.}
  \bibinfo{year}{2021}\natexlab{a}.
\newblock \showarticletitle{Improving code summarization with block-wise
  abstract syntax tree splitting}. In \bibinfo{booktitle}{\emph{2021 IEEE/ACM
  29th International Conference on Program Comprehension (ICPC)}}. IEEE,
  \bibinfo{pages}{184--195}.
\newblock


\bibitem[\protect\citeauthoryear{Lin}{Lin}{2004}]%
        {lin2004rouge}
\bibfield{author}{\bibinfo{person}{Chin-Yew Lin}.}
  \bibinfo{year}{2004}\natexlab{}.
\newblock \showarticletitle{Rouge: A package for automatic evaluation of
  summaries}. In \bibinfo{booktitle}{\emph{Text summarization branches out}}.
  \bibinfo{pages}{74--81}.
\newblock


\bibitem[\protect\citeauthoryear{Liu, Qiu, and Huang}{Liu
  et~al\mbox{.}}{2016}]%
        {liu2016recurrent}
\bibfield{author}{\bibinfo{person}{Pengfei Liu}, \bibinfo{person}{Xipeng Qiu},
  {and} \bibinfo{person}{Xuanjing Huang}.} \bibinfo{year}{2016}\natexlab{}.
\newblock \showarticletitle{Recurrent neural network for text classification
  with multi-task learning}.
\newblock \bibinfo{journal}{\emph{arXiv preprint arXiv:1605.05101}}
  (\bibinfo{year}{2016}).
\newblock


\bibitem[\protect\citeauthoryear{Mastropaolo, Cooper, Palacio, Scalabrino,
  Poshyvanyk, Oliveto, and Bavota}{Mastropaolo et~al\mbox{.}}{2022}]%
        {mastropaolo2022using}
\bibfield{author}{\bibinfo{person}{Antonio Mastropaolo},
  \bibinfo{person}{Nathan Cooper}, \bibinfo{person}{David~Nader Palacio},
  \bibinfo{person}{Simone Scalabrino}, \bibinfo{person}{Denys Poshyvanyk},
  \bibinfo{person}{Rocco Oliveto}, {and} \bibinfo{person}{Gabriele Bavota}.}
  \bibinfo{year}{2022}\natexlab{}.
\newblock \showarticletitle{Using Transfer Learning for Code-Related Tasks}.
\newblock \bibinfo{journal}{\emph{IEEE Transactions on Software Engineering}}
  (\bibinfo{year}{2022}).
\newblock


\bibitem[\protect\citeauthoryear{Min, Lyu, Holtzman, Artetxe, Lewis,
  Hajishirzi, and Zettlemoyer}{Min et~al\mbox{.}}{2022}]%
        {min2022rethinking}
\bibfield{author}{\bibinfo{person}{Sewon Min}, \bibinfo{person}{Xinxi Lyu},
  \bibinfo{person}{Ari Holtzman}, \bibinfo{person}{Mikel Artetxe},
  \bibinfo{person}{Mike Lewis}, \bibinfo{person}{Hannaneh Hajishirzi}, {and}
  \bibinfo{person}{Luke Zettlemoyer}.} \bibinfo{year}{2022}\natexlab{}.
\newblock \showarticletitle{Rethinking the Role of Demonstrations: What Makes
  In-Context Learning Work?}
\newblock \bibinfo{journal}{\emph{arXiv preprint arXiv:2202.12837}}
  (\bibinfo{year}{2022}).
\newblock


\bibitem[\protect\citeauthoryear{Mu, Chen, Shi, Wang, and Wang}{Mu
  et~al\mbox{.}}{2022}]%
        {mu2022automatic}
\bibfield{author}{\bibinfo{person}{Fangwen Mu}, \bibinfo{person}{Xiao Chen},
  \bibinfo{person}{Lin Shi}, \bibinfo{person}{Song Wang}, {and}
  \bibinfo{person}{Qing Wang}.} \bibinfo{year}{2022}\natexlab{}.
\newblock \showarticletitle{Automatic Comment Generation via Multi-Pass
  Deliberation}. In \bibinfo{booktitle}{\emph{37th IEEE/ACM International
  Conference on Automated Software Engineering}}. \bibinfo{pages}{1--12}.
\newblock


\bibitem[\protect\citeauthoryear{Mu, Chen, Shi, Wang, and Wang}{Mu
  et~al\mbox{.}}{2023}]%
        {mu2023developer}
\bibfield{author}{\bibinfo{person}{Fangwen Mu}, \bibinfo{person}{Xiao Chen},
  \bibinfo{person}{Lin Shi}, \bibinfo{person}{Song Wang}, {and}
  \bibinfo{person}{Qing Wang}.} \bibinfo{year}{2023}\natexlab{}.
\newblock \showarticletitle{Developer-Intent Driven Code Comment Generation}.
\newblock \bibinfo{journal}{\emph{arXiv preprint arXiv:2302.07055}}
  (\bibinfo{year}{2023}).
\newblock


\bibitem[\protect\citeauthoryear{Nashid, Sintaha, and Mesbah}{Nashid
  et~al\mbox{.}}{2023}]%
        {nashid2023retrieval}
\bibfield{author}{\bibinfo{person}{Noor Nashid}, \bibinfo{person}{Mifta
  Sintaha}, {and} \bibinfo{person}{Ali Mesbah}.}
  \bibinfo{year}{2023}\natexlab{}.
\newblock \showarticletitle{Retrieval-Based Prompt Selection for Code-Related
  Few-Shot Learning}. In \bibinfo{booktitle}{\emph{2023 IEEE/ACM 45th
  International Conference on Software Engineering (ICSE)}}. IEEE.
\newblock


\bibitem[\protect\citeauthoryear{Ni, Iyer, Radev, Stoyanov, Yih, Wang, and
  Lin}{Ni et~al\mbox{.}}{2023}]%
        {ni2023lever}
\bibfield{author}{\bibinfo{person}{Ansong Ni}, \bibinfo{person}{Srini Iyer},
  \bibinfo{person}{Dragomir Radev}, \bibinfo{person}{Ves Stoyanov},
  \bibinfo{person}{Wen-tau Yih}, \bibinfo{person}{Sida~I Wang}, {and}
  \bibinfo{person}{Xi~Victoria Lin}.} \bibinfo{year}{2023}\natexlab{}.
\newblock \showarticletitle{LEVER: Learning to Verify Language-to-Code
  Generation with Execution}.
\newblock \bibinfo{journal}{\emph{arXiv preprint arXiv:2302.08468}}
  (\bibinfo{year}{2023}).
\newblock


\bibitem[\protect\citeauthoryear{Niwattanakul, Singthongchai, Naenudorn, and
  Wanapu}{Niwattanakul et~al\mbox{.}}{2013}]%
        {niwattanakul2013using}
\bibfield{author}{\bibinfo{person}{Suphakit Niwattanakul},
  \bibinfo{person}{Jatsada Singthongchai}, \bibinfo{person}{Ekkachai
  Naenudorn}, {and} \bibinfo{person}{Supachanun Wanapu}.}
  \bibinfo{year}{2013}\natexlab{}.
\newblock \showarticletitle{Using of Jaccard coefficient for keywords
  similarity}. In \bibinfo{booktitle}{\emph{Proceedings of the international
  multiconference of engineers and computer scientists}},
  Vol.~\bibinfo{volume}{1}. \bibinfo{pages}{380--384}.
\newblock


\bibitem[\protect\citeauthoryear{Papineni, Roukos, Ward, and Zhu}{Papineni
  et~al\mbox{.}}{2002}]%
        {papineni2002bleu}
\bibfield{author}{\bibinfo{person}{Kishore Papineni}, \bibinfo{person}{Salim
  Roukos}, \bibinfo{person}{Todd Ward}, {and} \bibinfo{person}{Wei-Jing Zhu}.}
  \bibinfo{year}{2002}\natexlab{}.
\newblock \showarticletitle{Bleu: a method for automatic evaluation of machine
  translation}. In \bibinfo{booktitle}{\emph{Proceedings of the 40th annual
  meeting of the Association for Computational Linguistics}}.
  \bibinfo{pages}{311--318}.
\newblock


\bibitem[\protect\citeauthoryear{Pearce, Ahmad, Tan, Dolan-Gavitt, and
  Karri}{Pearce et~al\mbox{.}}{2022}]%
        {pearce2022asleep}
\bibfield{author}{\bibinfo{person}{Hammond Pearce}, \bibinfo{person}{Baleegh
  Ahmad}, \bibinfo{person}{Benjamin Tan}, \bibinfo{person}{Brendan
  Dolan-Gavitt}, {and} \bibinfo{person}{Ramesh Karri}.}
  \bibinfo{year}{2022}\natexlab{}.
\newblock \showarticletitle{Asleep at the keyboard? assessing the security of
  github copilot’s code contributions}. In \bibinfo{booktitle}{\emph{2022
  IEEE Symposium on Security and Privacy (SP)}}. IEEE,
  \bibinfo{pages}{754--768}.
\newblock


\bibitem[\protect\citeauthoryear{Pradel and Sen}{Pradel and Sen}{2018}]%
        {pradel2018deepbugs}
\bibfield{author}{\bibinfo{person}{Michael Pradel} {and}
  \bibinfo{person}{Koushik Sen}.} \bibinfo{year}{2018}\natexlab{}.
\newblock \showarticletitle{Deepbugs: A learning approach to name-based bug
  detection}.
\newblock \bibinfo{journal}{\emph{Proceedings of the ACM on Programming
  Languages}} \bibinfo{volume}{2}, \bibinfo{number}{OOPSLA}
  (\bibinfo{year}{2018}), \bibinfo{pages}{1--25}.
\newblock


\bibitem[\protect\citeauthoryear{Prenner, Babii, and Robbes}{Prenner
  et~al\mbox{.}}{2022}]%
        {prenner2022can}
\bibfield{author}{\bibinfo{person}{Julian~Aron Prenner}, \bibinfo{person}{Hlib
  Babii}, {and} \bibinfo{person}{Romain Robbes}.}
  \bibinfo{year}{2022}\natexlab{}.
\newblock \showarticletitle{Can OpenAI's codex fix bugs? an evaluation on
  QuixBugs}. In \bibinfo{booktitle}{\emph{Proceedings of the Third
  International Workshop on Automated Program Repair}}.
  \bibinfo{pages}{69--75}.
\newblock


\bibitem[\protect\citeauthoryear{Radford, Narasimhan, Salimans, and
  Sutskever}{Radford et~al\mbox{.}}{2018}]%
        {radford2018improving}
\bibfield{author}{\bibinfo{person}{Alec Radford}, \bibinfo{person}{Karthik
  Narasimhan}, \bibinfo{person}{Tim Salimans}, {and} \bibinfo{person}{Ilya
  Sutskever}.} \bibinfo{year}{2018}\natexlab{}.
\newblock \showarticletitle{Improving language understanding by generative
  pre-training}.
\newblock  (\bibinfo{year}{2018}).
\newblock


\bibitem[\protect\citeauthoryear{Raffel, Shazeer, Roberts, Lee, Narang, Matena,
  Zhou, Li, and Liu}{Raffel et~al\mbox{.}}{2020}]%
        {raffel2020exploring}
\bibfield{author}{\bibinfo{person}{Colin Raffel}, \bibinfo{person}{Noam
  Shazeer}, \bibinfo{person}{Adam Roberts}, \bibinfo{person}{Katherine Lee},
  \bibinfo{person}{Sharan Narang}, \bibinfo{person}{Michael Matena},
  \bibinfo{person}{Yanqi Zhou}, \bibinfo{person}{Wei Li}, {and}
  \bibinfo{person}{Peter~J Liu}.} \bibinfo{year}{2020}\natexlab{}.
\newblock \showarticletitle{Exploring the limits of transfer learning with a
  unified text-to-text transformer}.
\newblock \bibinfo{journal}{\emph{The Journal of Machine Learning Research}}
  \bibinfo{volume}{21}, \bibinfo{number}{1} (\bibinfo{year}{2020}),
  \bibinfo{pages}{5485--5551}.
\newblock


\bibitem[\protect\citeauthoryear{Reimers and Gurevych}{Reimers and
  Gurevych}{2019}]%
        {reimers2019sentence}
\bibfield{author}{\bibinfo{person}{Nils Reimers} {and} \bibinfo{person}{Iryna
  Gurevych}.} \bibinfo{year}{2019}\natexlab{}.
\newblock \showarticletitle{Sentence-bert: Sentence embeddings using siamese
  bert-networks}.
\newblock \bibinfo{journal}{\emph{arXiv preprint arXiv:1908.10084}}
  (\bibinfo{year}{2019}).
\newblock


\bibitem[\protect\citeauthoryear{Rodeghero, McMillan, McBurney, Bosch, and
  D'Mello}{Rodeghero et~al\mbox{.}}{2014}]%
        {rodeghero2014improving}
\bibfield{author}{\bibinfo{person}{Paige Rodeghero}, \bibinfo{person}{Collin
  McMillan}, \bibinfo{person}{Paul~W McBurney}, \bibinfo{person}{Nigel Bosch},
  {and} \bibinfo{person}{Sidney D'Mello}.} \bibinfo{year}{2014}\natexlab{}.
\newblock \showarticletitle{Improving automated source code summarization via
  an eye-tracking study of programmers}. In
  \bibinfo{booktitle}{\emph{Proceedings of the 36th international conference on
  Software engineering}}. \bibinfo{pages}{390--401}.
\newblock


\bibitem[\protect\citeauthoryear{Roy, Fakhoury, and Arnaoudova}{Roy
  et~al\mbox{.}}{2021}]%
        {roy2021reassessing}
\bibfield{author}{\bibinfo{person}{Devjeet Roy}, \bibinfo{person}{Sarah
  Fakhoury}, {and} \bibinfo{person}{Venera Arnaoudova}.}
  \bibinfo{year}{2021}\natexlab{}.
\newblock \showarticletitle{Reassessing automatic evaluation metrics for code
  summarization tasks}. In \bibinfo{booktitle}{\emph{Proceedings of the 29th
  ACM Joint Meeting on European Software Engineering Conference and Symposium
  on the Foundations of Software Engineering}}. \bibinfo{pages}{1105--1116}.
\newblock


\bibitem[\protect\citeauthoryear{Rubin, Herzig, and Berant}{Rubin
  et~al\mbox{.}}{2021}]%
        {rubin2021learning}
\bibfield{author}{\bibinfo{person}{Ohad Rubin}, \bibinfo{person}{Jonathan
  Herzig}, {and} \bibinfo{person}{Jonathan Berant}.}
  \bibinfo{year}{2021}\natexlab{}.
\newblock \showarticletitle{Learning to retrieve prompts for in-context
  learning}.
\newblock \bibinfo{journal}{\emph{arXiv preprint arXiv:2112.08633}}
  (\bibinfo{year}{2021}).
\newblock


\bibitem[\protect\citeauthoryear{Shi, Fried, Ghazvininejad, Zettlemoyer, and
  Wang}{Shi et~al\mbox{.}}{2022}]%
        {shi2022natural}
\bibfield{author}{\bibinfo{person}{Freda Shi}, \bibinfo{person}{Daniel Fried},
  \bibinfo{person}{Marjan Ghazvininejad}, \bibinfo{person}{Luke Zettlemoyer},
  {and} \bibinfo{person}{Sida~I Wang}.} \bibinfo{year}{2022}\natexlab{}.
\newblock \showarticletitle{Natural language to code translation with
  execution}.
\newblock \bibinfo{journal}{\emph{arXiv preprint arXiv:2204.11454}}
  (\bibinfo{year}{2022}).
\newblock


\bibitem[\protect\citeauthoryear{Wan, Zhao, Yang, Xu, Ying, Wu, and Yu}{Wan
  et~al\mbox{.}}{2018}]%
        {wan2018improving}
\bibfield{author}{\bibinfo{person}{Yao Wan}, \bibinfo{person}{Zhou Zhao},
  \bibinfo{person}{Min Yang}, \bibinfo{person}{Guandong Xu},
  \bibinfo{person}{Haochao Ying}, \bibinfo{person}{Jian Wu}, {and}
  \bibinfo{person}{Philip~S Yu}.} \bibinfo{year}{2018}\natexlab{}.
\newblock \showarticletitle{Improving automatic source code summarization via
  deep reinforcement learning}. In \bibinfo{booktitle}{\emph{Proceedings of the
  33rd ACM/IEEE international conference on automated software engineering}}.
  \bibinfo{pages}{397--407}.
\newblock


\bibitem[\protect\citeauthoryear{Wang, Yang, Gao, Peng, Zhang, and Lyu}{Wang
  et~al\mbox{.}}{2022}]%
        {wang2022no}
\bibfield{author}{\bibinfo{person}{Chaozheng Wang}, \bibinfo{person}{Yuanhang
  Yang}, \bibinfo{person}{Cuiyun Gao}, \bibinfo{person}{Yun Peng},
  \bibinfo{person}{Hongyu Zhang}, {and} \bibinfo{person}{Michael~R. Lyu}.}
  \bibinfo{year}{2022}\natexlab{}.
\newblock \showarticletitle{No More Fine-Tuning? An Experimental Evaluation of
  Prompt Tuning in Code Intelligence}. In \bibinfo{booktitle}{\emph{Proceedings
  of the 30th ACM Joint European Software Engineering Conference and Symposium
  on the Foundations of Software Engineering}} (Singapore, Singapore)
  \emph{(\bibinfo{series}{ESEC/FSE 2022})}. \bibinfo{publisher}{Association for
  Computing Machinery}, \bibinfo{address}{New York, NY, USA},
  \bibinfo{pages}{382–394}.
\newblock
\urldef\tempurl%
\url{https://doi.org/10.1145/3540250.3549113}
\showDOI{\tempurl}


\bibitem[\protect\citeauthoryear{Wang, Li, Ma, Xia, and Jin}{Wang
  et~al\mbox{.}}{2020a}]%
        {wang2020detecting}
\bibfield{author}{\bibinfo{person}{Wenhan Wang}, \bibinfo{person}{Ge Li},
  \bibinfo{person}{Bo Ma}, \bibinfo{person}{Xin Xia}, {and}
  \bibinfo{person}{Zhi Jin}.} \bibinfo{year}{2020}\natexlab{a}.
\newblock \showarticletitle{Detecting code clones with graph neural network and
  flow-augmented abstract syntax tree}. In \bibinfo{booktitle}{\emph{2020 IEEE
  27th International Conference on Software Analysis, Evolution and
  Reengineering (SANER)}}. IEEE, \bibinfo{pages}{261--271}.
\newblock


\bibitem[\protect\citeauthoryear{Wang, Zhang, Sui, Wan, Zhao, Wu, Philip, and
  Xu}{Wang et~al\mbox{.}}{2020b}]%
        {wang2020reinforcement}
\bibfield{author}{\bibinfo{person}{Wenhua Wang}, \bibinfo{person}{Yuqun Zhang},
  \bibinfo{person}{Yulei Sui}, \bibinfo{person}{Yao Wan}, \bibinfo{person}{Zhou
  Zhao}, \bibinfo{person}{Jian Wu}, \bibinfo{person}{S~Yu Philip}, {and}
  \bibinfo{person}{Guandong Xu}.} \bibinfo{year}{2020}\natexlab{b}.
\newblock \showarticletitle{Reinforcement-learning-guided source code
  summarization using hierarchical attention}.
\newblock \bibinfo{journal}{\emph{IEEE Transactions on software Engineering}}
  \bibinfo{volume}{48}, \bibinfo{number}{1} (\bibinfo{year}{2020}),
  \bibinfo{pages}{102--119}.
\newblock


\bibitem[\protect\citeauthoryear{Wang, Shi, Du, Yang, Hu, Han, Zhang, and
  Zhang}{Wang et~al\mbox{.}}{2021a}]%
        {wang2021cocosum}
\bibfield{author}{\bibinfo{person}{Yanlin Wang}, \bibinfo{person}{Ensheng Shi},
  \bibinfo{person}{Lun Du}, \bibinfo{person}{Xiaodi Yang},
  \bibinfo{person}{Yuxuan Hu}, \bibinfo{person}{Shi Han},
  \bibinfo{person}{Hongyu Zhang}, {and} \bibinfo{person}{Dongmei Zhang}.}
  \bibinfo{year}{2021}\natexlab{a}.
\newblock \showarticletitle{Cocosum: Contextual code summarization with
  multi-relational graph neural network}.
\newblock \bibinfo{journal}{\emph{arXiv preprint arXiv:2107.01933}}
  (\bibinfo{year}{2021}).
\newblock


\bibitem[\protect\citeauthoryear{Wang, Wang, Joty, and Hoi}{Wang
  et~al\mbox{.}}{2021b}]%
        {wang2021codet5}
\bibfield{author}{\bibinfo{person}{Yue Wang}, \bibinfo{person}{Weishi Wang},
  \bibinfo{person}{Shafiq Joty}, {and} \bibinfo{person}{Steven~CH Hoi}.}
  \bibinfo{year}{2021}\natexlab{b}.
\newblock \showarticletitle{Codet5: Identifier-aware unified pre-trained
  encoder-decoder models for code understanding and generation}.
\newblock \bibinfo{journal}{\emph{arXiv preprint arXiv:2109.00859}}
  (\bibinfo{year}{2021}).
\newblock


\bibitem[\protect\citeauthoryear{Wei, Li, Xia, Fu, and Jin}{Wei
  et~al\mbox{.}}{2019}]%
        {wei2019code}
\bibfield{author}{\bibinfo{person}{Bolin Wei}, \bibinfo{person}{Ge Li},
  \bibinfo{person}{Xin Xia}, \bibinfo{person}{Zhiyi Fu}, {and}
  \bibinfo{person}{Zhi Jin}.} \bibinfo{year}{2019}\natexlab{}.
\newblock \showarticletitle{Code generation as a dual task of code
  summarization}.
\newblock \bibinfo{journal}{\emph{Advances in neural information processing
  systems}}  \bibinfo{volume}{32} (\bibinfo{year}{2019}).
\newblock


\bibitem[\protect\citeauthoryear{Wei, Li, Li, Xia, and Jin}{Wei
  et~al\mbox{.}}{2020}]%
        {wei2020retrieve}
\bibfield{author}{\bibinfo{person}{Bolin Wei}, \bibinfo{person}{Yongmin Li},
  \bibinfo{person}{Ge Li}, \bibinfo{person}{Xin Xia}, {and}
  \bibinfo{person}{Zhi Jin}.} \bibinfo{year}{2020}\natexlab{}.
\newblock \showarticletitle{Retrieve and refine: exemplar-based neural comment
  generation}. In \bibinfo{booktitle}{\emph{Proceedings of the 35th IEEE/ACM
  International Conference on Automated Software Engineering}}.
  \bibinfo{pages}{349--360}.
\newblock


\bibitem[\protect\citeauthoryear{Wong, Liu, and Tan}{Wong
  et~al\mbox{.}}{2015}]%
        {wong2015clocom}
\bibfield{author}{\bibinfo{person}{Edmund Wong}, \bibinfo{person}{Taiyue Liu},
  {and} \bibinfo{person}{Lin Tan}.} \bibinfo{year}{2015}\natexlab{}.
\newblock \showarticletitle{Clocom: Mining existing source code for automatic
  comment generation}. In \bibinfo{booktitle}{\emph{2015 IEEE 22nd
  International Conference on Software Analysis, Evolution, and Reengineering
  (SANER)}}. IEEE, \bibinfo{pages}{380--389}.
\newblock


\bibitem[\protect\citeauthoryear{Wong, Yang, and Tan}{Wong
  et~al\mbox{.}}{2013}]%
        {wong2013autocomment}
\bibfield{author}{\bibinfo{person}{Edmund Wong}, \bibinfo{person}{Jinqiu Yang},
  {and} \bibinfo{person}{Lin Tan}.} \bibinfo{year}{2013}\natexlab{}.
\newblock \showarticletitle{Autocomment: Mining question and answer sites for
  automatic comment generation}. In \bibinfo{booktitle}{\emph{2013 28th
  IEEE/ACM International Conference on Automated Software Engineering (ASE)}}.
  IEEE, \bibinfo{pages}{562--567}.
\newblock


\bibitem[\protect\citeauthoryear{Xie, Ye, Sun, and Zhang}{Xie
  et~al\mbox{.}}{2021}]%
        {xie2021exploiting}
\bibfield{author}{\bibinfo{person}{Rui Xie}, \bibinfo{person}{Wei Ye},
  \bibinfo{person}{Jinan Sun}, {and} \bibinfo{person}{Shikun Zhang}.}
  \bibinfo{year}{2021}\natexlab{}.
\newblock \showarticletitle{Exploiting method names to improve code
  summarization: A deliberation multi-task learning approach}. In
  \bibinfo{booktitle}{\emph{2021 IEEE/ACM 29th International Conference on
  Program Comprehension (ICPC)}}. IEEE, \bibinfo{pages}{138--148}.
\newblock


\bibitem[\protect\citeauthoryear{Ye, Xie, Zhang, Hu, Wang, and Zhang}{Ye
  et~al\mbox{.}}{2020}]%
        {ye2020leveraging}
\bibfield{author}{\bibinfo{person}{Wei Ye}, \bibinfo{person}{Rui Xie},
  \bibinfo{person}{Jinglei Zhang}, \bibinfo{person}{Tianxiang Hu},
  \bibinfo{person}{Xiaoyin Wang}, {and} \bibinfo{person}{Shikun Zhang}.}
  \bibinfo{year}{2020}\natexlab{}.
\newblock \showarticletitle{Leveraging code generation to improve code
  retrieval and summarization via dual learning}. In
  \bibinfo{booktitle}{\emph{Proceedings of The Web Conference 2020}}.
  \bibinfo{pages}{2309--2319}.
\newblock


\bibitem[\protect\citeauthoryear{Zeng, Yu, Li, Xia, Wang, Geng, Bai, Dong, and
  Liao}{Zeng et~al\mbox{.}}{2022}]%
        {zeng2022degraphcs}
\bibfield{author}{\bibinfo{person}{Chen Zeng}, \bibinfo{person}{Yue Yu},
  \bibinfo{person}{Shanshan Li}, \bibinfo{person}{Xin Xia},
  \bibinfo{person}{Zhiming Wang}, \bibinfo{person}{Mingyang Geng},
  \bibinfo{person}{Linxiao Bai}, \bibinfo{person}{Wei Dong}, {and}
  \bibinfo{person}{Xiangke Liao}.} \bibinfo{year}{2022}\natexlab{}.
\newblock \showarticletitle{deGraphCS: Embedding Variable-based Flow Graph for
  Neural Code Search}.
\newblock \bibinfo{journal}{\emph{ACM Transactions on Software Engineering and
  Methodology (TOSEM)}} (\bibinfo{year}{2022}).
\newblock


\bibitem[\protect\citeauthoryear{Zhai, Xu, Shi, Tao, Pan, Ma, Xu, Zhang, Tan,
  and Zhang}{Zhai et~al\mbox{.}}{2020}]%
        {zhai2020cpc}
\bibfield{author}{\bibinfo{person}{Juan Zhai}, \bibinfo{person}{Xiangzhe Xu},
  \bibinfo{person}{Yu Shi}, \bibinfo{person}{Guanhong Tao},
  \bibinfo{person}{Minxue Pan}, \bibinfo{person}{Shiqing Ma},
  \bibinfo{person}{Lei Xu}, \bibinfo{person}{Weifeng Zhang},
  \bibinfo{person}{Lin Tan}, {and} \bibinfo{person}{Xiangyu Zhang}.}
  \bibinfo{year}{2020}\natexlab{}.
\newblock \showarticletitle{CPC: Automatically classifying and propagating
  natural language comments via program analysis}. In
  \bibinfo{booktitle}{\emph{Proceedings of the ACM/IEEE 42nd International
  Conference on Software Engineering}}. \bibinfo{pages}{1359--1371}.
\newblock


\bibitem[\protect\citeauthoryear{Zhang, Fang, Ge, Li, and Liu}{Zhang
  et~al\mbox{.}}{2023}]%
        {zhang2023efficient}
\bibfield{author}{\bibinfo{person}{Aiping Zhang}, \bibinfo{person}{Liming
  Fang}, \bibinfo{person}{Chunpeng Ge}, \bibinfo{person}{Piji Li}, {and}
  \bibinfo{person}{Zhe Liu}.} \bibinfo{year}{2023}\natexlab{}.
\newblock \showarticletitle{Efficient transformer with code token learner for
  code clone detection}.
\newblock \bibinfo{journal}{\emph{Journal of Systems and Software}}
  \bibinfo{volume}{197} (\bibinfo{year}{2023}), \bibinfo{pages}{111557}.
\newblock


\bibitem[\protect\citeauthoryear{Zhang, Wang, Zhang, Sun, and Liu}{Zhang
  et~al\mbox{.}}{2020}]%
        {zhang2020retrieval}
\bibfield{author}{\bibinfo{person}{Jian Zhang}, \bibinfo{person}{Xu Wang},
  \bibinfo{person}{Hongyu Zhang}, \bibinfo{person}{Hailong Sun}, {and}
  \bibinfo{person}{Xudong Liu}.} \bibinfo{year}{2020}\natexlab{}.
\newblock \showarticletitle{Retrieval-based neural source code summarization}.
  In \bibinfo{booktitle}{\emph{Proceedings of the ACM/IEEE 42nd International
  Conference on Software Engineering}}. \bibinfo{pages}{1385--1397}.
\newblock


\bibitem[\protect\citeauthoryear{Zhang, Yu, Hashimoto, Lewis, Yih, Fried, and
  Wang}{Zhang et~al\mbox{.}}{2022}]%
        {zhang2022coder}
\bibfield{author}{\bibinfo{person}{Tianyi Zhang}, \bibinfo{person}{Tao Yu},
  \bibinfo{person}{Tatsunori~B Hashimoto}, \bibinfo{person}{Mike Lewis},
  \bibinfo{person}{Wen-tau Yih}, \bibinfo{person}{Daniel Fried}, {and}
  \bibinfo{person}{Sida~I Wang}.} \bibinfo{year}{2022}\natexlab{}.
\newblock \showarticletitle{Coder Reviewer Reranking for Code Generation}.
\newblock \bibinfo{journal}{\emph{arXiv preprint arXiv:2211.16490}}
  (\bibinfo{year}{2022}).
\newblock


\end{thebibliography}
